\documentclass[12pt]{iopart}

\usepackage{iopams}  
\usepackage{color}
\usepackage[dvipsnames]{xcolor} 
\usepackage{tikz}
\usepackage{psfrag}
\usepackage{graphicx}
\usepackage{stackrel}
\usepackage{amsfonts}
\usepackage{bm}          
\usepackage{amssymb}
\usepackage{perpage}
\MakePerPage{footnote}

\newcommand{\be}{\begin{equation}}
\newcommand{\ee}{\end{equation}}
\newcommand{\ba}{\begin{eqnarray}}
\newcommand{\ea}{\end{eqnarray}}

\begin{document}

\title[]{Transfer matrix computation of generalised critical polynomials in percolation}

\author{Christian R.\ Scullard$^{1}$ and Jesper Lykke Jacobsen$^{2,3}$}
\address{${}^1$Lawrence Livermore National Laboratory, Livermore CA 94550, USA}
\address{${}^2$LPTENS, \'Ecole Normale Sup\'erieure, 24 rue Lhomond, 75231 Paris, France}
\address{${}^3$Universit\'e Pierre et Marie Curie, 4 place Jussieu, 75252 Paris, France}

\eads{\mailto{scullard1@llnl.gov},\mailto{jesper.jacobsen@ens.fr}}

\begin{abstract}

  Percolation thresholds have recently been studied by means of a
  graph polynomial $P_B(p)$, henceforth referred to as the critical
  polynomial, that may be defined on any periodic lattice. The
  polynomial depends on a finite subgraph $B$, called the basis, and
  the way in which the basis is tiled to form the lattice. The unique
  root of $P_B(p)$ in $[0,1]$ either gives the exact percolation
  threshold for the lattice, or provides an approximation that becomes
  more accurate with appropriately increasing size of $B$. Initially
  $P_B(p)$ was defined by a contraction-deletion identity, similar to
  that satisfied by the Tutte polynomial.  Here, we give an
  alternative probabilistic definition of $P_B(p)$, which allows for
  much more efficient computations, by using the transfer matrix, than
  was previously possible with contraction-deletion.

  We present bond percolation polynomials for the $(4,8^2)$, kagome, 
  and $(3,12^2)$ lattices for bases of up to respectively 96, 162, and
  243 edges, much larger than the previous limit of 36 edges using
  contraction-deletion. We discuss in detail the role of the
  symmetries and the embedding of $B$.  For the largest bases, we
  obtain the thresholds $p_c(4,8^2) = 0.676\,803\,329 \cdots$, $p_c(\mathrm{kagome}) = 0.524\,404\,998
  \cdots$, $p_c(3,12^2) = 0.740\,420\,798 \cdots$, comparable to the best simulation results.
  We also show that the alternative definition of $P_B(p)$ can be
  applied to study site percolation problems.

\end{abstract}

\noindent

\section{Introduction}

Since its introduction \cite{BroadbentHammersley57}, percolation has provided a wealth of problems
for physicists, mathematicians, and computer scientists. One of the
most difficult is the analytical determination of critical
probabilities. Given an infinite $d-$dimensional lattice $L$, declare each edge of $L$ to
be open with probability $p$ and closed with probability
$1-p$. Between the regimes of sparse clusters near $p=0$ and the
nearly complete filling of space near $p=1$ lies the critical
probability (also called the {\em percolation threshold}), $p_c$, below
which all clusters are finite but above which there is an infinite
cluster. In site percolation, which we will also consider here, the
vertices of the graph are occupied or unoccupied with probability $p$
or $1-p$, and percolation clusters can be defined by declaring an edge
open when it connects two occupied vertices.

For $d=1$, percolation is trivial and $p_c=1$, but for $d \ge 3$, the
problem is completely unsolved. In two dimensions, bond and site probabilities can be found only on a narrow class of lattices
formed from self-dual 3-uniform hypergraphs. In these cases the
threshold is given as the root in $[0,1]$ of a finite
polynomial. Previously, it was shown \cite{ScullardZiff10,Scullard11,Scullard12} that the
concept of a critical polynomial may be extended to {\it any}
two-dimensional lattice. The unique root in $[0,1]$ of this polynomial
provides the exact $p_c$ for
lattices in the solvable class, and for unsolved problems, where we call it the {\em generalised} critical polynomial, it gives
answers that can seemingly be brought arbitrarily close to the exact
threshold. Recently, we extended the definition of
this polynomial to the full $q$-state Potts model and used it to
explore the phase diagram of the kagome lattice \cite{Jacobsen12}.

\begin{figure}
\begin{center}
\includegraphics{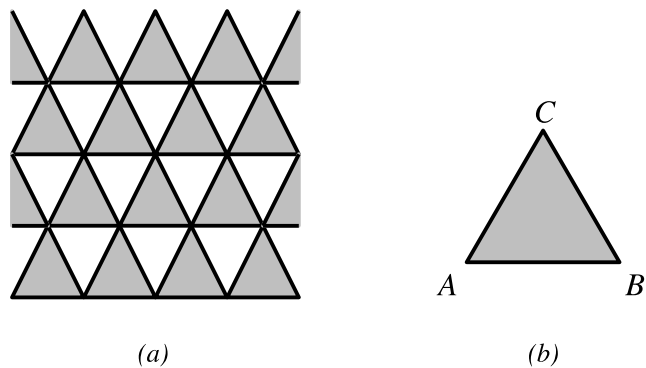}
\caption{a) A class of self-dual 3-uniform hypergraphs. b) Each shaded
  triangle may contain arbitrary interactions among its boundary
  vertices $A$, $B$ and $C$.}
\label{fig:selfdual}
\end{center}
\end{figure}

\section{The generalised critical polynomial}

We first consider bond percolation on the self-dual 3-uniform
hypergraph depicted in Figure~\ref{fig:selfdual}a. The particular
hypergraph shown is of the simple triangular type, but the argument
can be extended to other types of self-dual 3-uniform hypergraphs
\cite{ZiffScullard06}; one can also treat site percolation problems by
reasoning on the covering lattice \cite{Ziff06} or by introducting correlations \cite{Scullard06}. Interior to the
boundary vertices of each shaded triangle (Figure~\ref{fig:selfdual}b),
we may have essentially any network of bonds, correlations, and
sites. The critical point of such a system is given by
\cite{Ziff06,BollobasRiordan11}
\begin{equation}
 P(A,B,C)=P(\bar{A},\bar{B},\bar{C}) \,, \label{eq:allnone}
\end{equation}
where $P(A,B,C)$ is the probability that all three boundary vertices
are connected by an open path in the triangle, and
$P(\bar{A},\bar{B},\bar{C})$ is the probability that none are
connected. The result of applying this condition is a polynomial in
the probability $p$ with degree equal to the number of randomly
occupied elements (edges for bond percolation, or vertices for site
percolation) within a triangle.  Thus, all thresholds that are known
exactly are algebraic numbers.  We may also consider inhomogeneous
percolation in which each edge $i$ is assigned a different probability
$p_i$ so that (\ref{eq:allnone}) provides a critical surface within the
space of all $p_i$'s.

As already mentioned, a critical polynomial $P_B(p)$ can be defined
more generally for bond percolation on any two-dimensional lattice
\cite{ScullardZiff10,Scullard11,Scullard12,Scullard11-2}.  It depends
on a finite subgraph $B$, called the basis, and its embedding into the
infinite lattice $L$. This $P_B(p)$ indeed reproduces the exact
percolation threshold (\ref{eq:allnone}) for exactly solvable cases,
but in general it is only an approximation that however converges very
rapidly to the true $p_c$ upon appropriately increasing the size of
$B$. The definition of $P_B(p)$ used in these works proceeds by
applying a contraction-deletion principle to the edges in $B$, and by
this fact it can be further generalised \cite{Jacobsen12} to a critical
polynomial $P_B(q,v)$ for the $q$-state Potts model with temperature
parameter $v$.

\begin{figure}
\begin{center}
\includegraphics{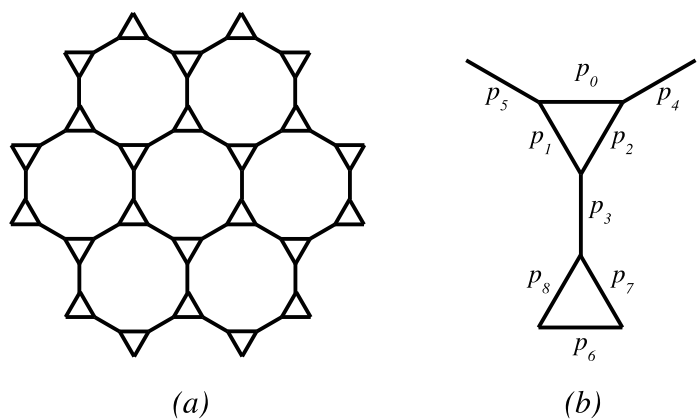}
\caption{a) The $(3,12^2)$ lattice; b) the assignment of probabilities on the unit cell.}
\label{fig:3-12lattice}
\end{center}
\end{figure}

We recall here the contraction-deletion definition of $P_B(p)$ by
means of a specific example.  Consider for $L$ the $(3,12^2)$ lattice,
shown in Figure~\ref{fig:3-12lattice}a. Its threshold is not known
exactly, but has been the subject of much numerical
\cite{Parviainen,ZiffGu09,Ding10} and analytical \cite{ZiffGu09,ScullardZiff06,MayWierman06,Tsallis}
work. For the basis $B$ we choose the unit cell shown in
Figure~\ref{fig:3-12lattice}b with an arbitrary inhomogeneous
assignment of probabilities $p_i$ to the nine edges. Notice that $B$
is embedded into $L$ in a checkerboard fashion. Any edge
of $L$ is a translation of an edge in $B$ and is therefore assigned
the corresponding probability $p_i$ for some $i=0,1,\ldots,8$.

\begin{figure}
\begin{center}
\includegraphics{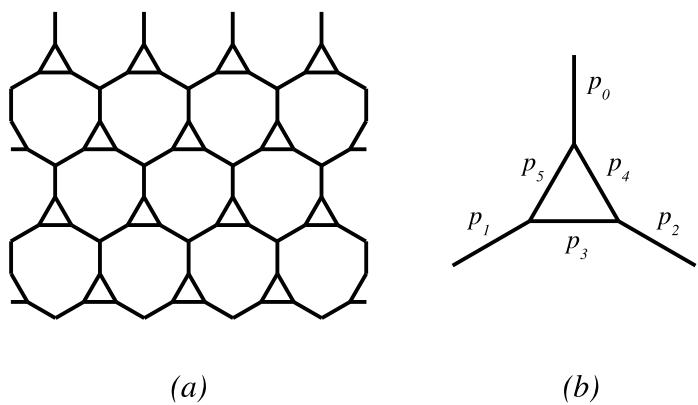}
\caption{a) The martini lattice; b) the assignment of probabilities on the unit cell.}
\label{fig:martini}
\end{center}
\end{figure}

\begin{figure}
\begin{center}
\includegraphics{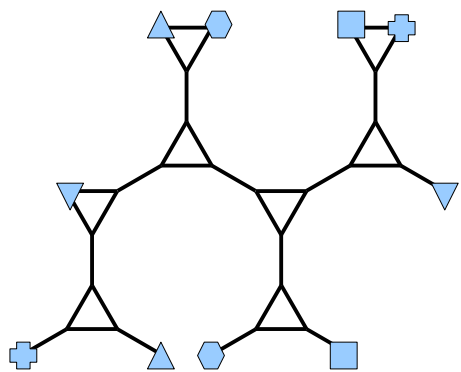}
\caption{A $36-$edge basis for the $(3,12^2)$ lattice. Each edge should be understood to have a different probability, and the shapes on the terminals indicate how this basis is embedded into the lattice.}
\label{fig:3-12}
\end{center}
\end{figure}

If we {\em delete} the
$p_0$ edge by setting $p_0=0$ in Figure~\ref{fig:3-12lattice}b, we
obtain the martini lattice (Figure~\ref{fig:martini}a) with some edges
coupled in series. Similarly, we can {\em contract} the $p_0$ edge by
setting $p_0=1$, and we again find the martini lattice, but with some
edges coupled in series and parallel.  In both cases, the coupled
edges can be replaced by simple edges with appropriate effective
percolation probabilities.  These considerations lead to the following
expression for the critical surface $P_B = TT$ of the $(3,12^2)$
lattice:
\begin{eqnarray}
TT(p_0,p_1,\ldots,p_8)&=&p_0 M(p_3 [p_1+p_2-p_1 p_2],p_4,p_5,p_6,p_7,p_8) \cr
&+& (1-p_0) M(p_3,p_2 p_4,p_1 p_5,p_6,p_7,p_8) \,,
\end{eqnarray}
where $M$ is the corresponding expression for the martini lattice
(Figure~\ref{fig:martini}a) with the inhomogeneous assignment of
probabilities to the basis shown in Figure~\ref{fig:martini}b.
However, the critical surface of the martini lattice can be found
exactly with (\ref{eq:allnone}), and inserting this we obtain finally
in the homogeneous case the critical polynomial
\begin{equation}
 TT(p,p,\ldots,p)=1 - 3 p^4 - 6 p^5 + 3 p^6 + 15 p^7 - 15 p^8 + 4 p^9 \,.
\end{equation}
The corresponding approximation to the percolation threshold reads
$TT(p,p,\ldots,p) = 0$, and its unique solution on $[0,1]$ is $p_c
= 0.740\,423\,31 \cdots$. Comparing this with the most accurately known
numerical value, $p_c^{\mathrm{num}}=0.740\,420\,77(2)$ \cite{Ding10},
we infer that the prediction provided by the $9^{\mathrm{th}}$-order
critical polynomial is close, but not exactly equal to the true $p_c$. However, the
approximation can be improved by increasing the size of the basis.
For example, using the basis of Figure~\ref{fig:3-12}, we find a
$36^{\mathrm{th}}$-order polynomial, reported in \cite{Scullard12},
that makes the prediction $p_c = 0.740\,420\,99 \cdots$, which is
closer to the numerical value.


Critical polynomials $P_B(p)$ defined in this way are unique, that is,
they are a property only of the basis $B$ and the way in which $B$ is
embedded in the infinite lattice $L$. In particular, $P_B(p)$ is
independent of the order in which edges are contracted-deleted. An
important property of $P_B(p)$ is that in all exactly solvable cases,
the smallest possible basis already provides the exact answer
(\ref{eq:allnone}), and the same answer invariably factorises from
$P_B(p)$ upon using a larger basis.  On the other hand, for unsolved
cases, using appropriate larger bases leads to predictions that
improve with the size of $B$, and appear to approach the true
$p_c$. How close one can get to $p_c$ is limited by one's ability to
actually compute the polynomial on large $B$. In \cite{Scullard12}, a
computer program was used to perform the contraction-deletion
algorithm on various bases for the Archimedean lattices. However, this
algorithm is exponential in the number of edges in $B$, and the upper
limit of feasibility was $36$ edges. Nevertheless, the corresponding
$P_B(p)$ yielded bond percolation thresholds that were generally
within $10^{-7}$ of the numerically determined values.

Below, we present an alternative definition of $P_B(p)$ in terms of
probabilities of events on $B$. This permits a much more efficient
calculation using a transfer matrix approach, where roughly speaking
the algorithm is exponential only in the number of vertices across a
horizontal cross-section of $B$.  In practice, this permits us to
compute the critical polynomial for bond percolation on the kagome and
$(4,8^2)$ lattices up to 162 and 96 edges respectively, and up to 243 edges for the
$(3,12^2)$ lattice.  The alternative definition also makes it possible
to address {\em site} percolation, and we present results for the
square and hexagonal lattices.
\begin{figure}
\begin{center}
\includegraphics{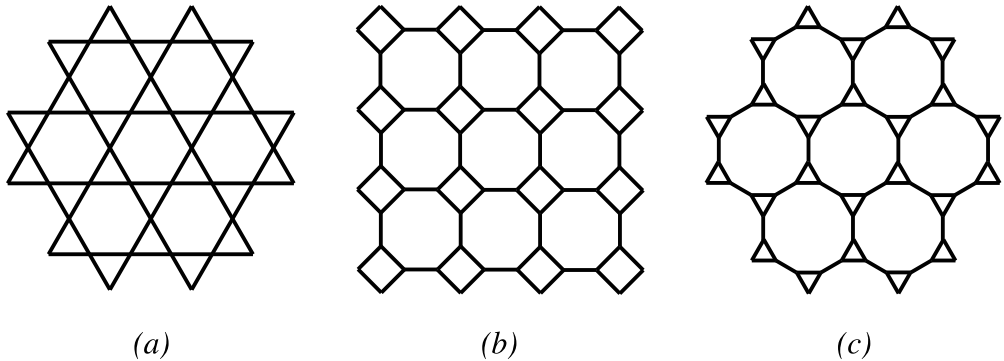}
\caption{a) the kagome lattice; b) the $(4,8^2)$ lattice; c) the $(3,12^2)$ lattice.}
\label{fig:bondlattices}
\end{center}
\end{figure}

\subsection{Alternative definition}
\label{sec:altdef}

In bond percolation, the probability of any event on the finite graph
$B$ is a sum of terms of the type $\prod_{i \in A} p_i \times \prod_{i
  \notin A} (1-p_i)$, where $A$ are some subsets of the edges in $B$
describing which edges need to be open in order to realise the
event. But if all factors $(1-p_i)$ are expanded out, one obtains
instead a sum of terms of the type $\prod_{i \in A'} p_i$, from which
it is in general difficult to deduce the subsets $A$ that provided the
geometrical characterisation of the event. The remedy is to
define $v_i = \frac{p_i}{1-p_i}$ so that, after multiplication with an
appropriate normalisation factor, the probabilities $p_i$ and
$(1-p_i)$ get replaced by $v_i$ and $1$. Any term of the type
$\prod_{i \in A} v_i$ then directly permits one to infer the
corresponding subset $A$ of open edges.

We are here interested in the probabilistic, geometrical
interpretation of the critical polynomials $P_B(p)$. But to discuss
this, we will first need some definitions.

The infinite lattice $L$ is partitioned into identical subgraphs $B$,
and we assume that each is in the same edge-state (or vertex-state for
site percolation). We are interested in the global connectivity
properties of the system. If, given any two copies of the basis, $B_1$
and $B_2$, separated by an arbitrary distance, it is possible to
travel from $B_1$ to $B_2$ along an open path, then we say that there
is an infinite two-dimensional (2D) cluster in the system. We denote
the probability of this event $P(2D;B)$. On the other hand, if it is
not possible to connect any non-neighbouring $B_1$ and $B_2$, then
there are no infinite clusters in the system, a situation whose
probability we write as $P(0D;B)$. The third possibility is that {\it
  some} arbitrarily separated $B_1$ and $B_2$ are connected, but not
all, indicating the presence of infinite one-dimensional (1D) paths
(or filaments), and we denote the corresponding probability
$P(1D;B)$. By normalisation of probabilities we obviously have
\begin{equation}
 P(0D;B)+P(1D;B)+P(2D;B)=1 \,.
\end{equation}
We have found that all the (inhomogeneous) critical polynomials
$P_B(\{p_i\})$ that we have computed
\cite{ScullardZiff10,Scullard11,Scullard12,Jacobsen12,Scullard11-2}
using the contraction-deletion definition can be rewritten very simply as
\begin{equation}
 P(2D;B)=P(0D;B) \,. \label{eq:2D0D}
\end{equation}
Despite its apparent simplicity, eq.~(\ref{eq:2D0D}) is the main
result of this paper. We leave it as an open problem to prove
mathematically that the probabilistic formula (\ref{eq:2D0D}) and
contraction-deletion both define the same polynomial $P_B(p)$ for {\em
  any} lattice $L$ and basis $B$. But in view of the circumstantial evidence
from the many examples that we have worked out using both definitions, we
shall henceforth suppose that they are indeed equivalent in general.

We further notice that (\ref{eq:2D0D}) has a number of pleasing
properties. First, it becomes (\ref{eq:allnone}) for the solvable
class of lattices, which is obviously the most basic
requirement. Second, it respects duality. Consider bond percolation on
the dual lattice $L_d$ in which we now study events that take place on
closed edges with probability $1-p$, a measure we denote $Q^*$. Then
it is clear that we have $Q^*(2D;B)=P(0D;B)$ and $Q^*(0D;B)=P(2D;B)$,
and thus the condition (\ref{eq:2D0D}) can be written in a variety of
forms,
\begin{eqnarray}
 P(2D;B)&=&Q^*(2D;B) \label{eq:PQ*} \\
 Q^*(0D;B)&=&P(0D;B) \\
 Q^*(0D;B)&=&Q^*(2D;B) \,.
\end{eqnarray}
This last equation indicates that our criterion may be applied to
closed bonds on $L_d$, with the result that the roots of $P_B(p)$
satisfy $p_c(L)=1-p_c(L_d)$, as required by duality.

The reason that (\ref{eq:allnone}) is the critical point of certain
lattices, is that it locates the probability at which the measure of
open paths is identical to that of closed paths on the dual. That this
implies criticality was assumed to be true at least since the work of
Sykes and Essam \cite{SykesEssam} in the 1960s, but has now been
rigourously established \cite{BollobasRiordan11}. For general
lattices, this self-dual point does not exist. Nevertheless,
universality asserts that equation (\ref{eq:PQ*}) should give
estimates of $p_c$ that become exact in the limit of infinite $B$. The
crossing probability $P(2D)$ exists in the scaling limit, and has been
studied in great detail in the conformal field theory literature
\cite{diFrancescoSaleurZuber1987,Pinson94,Ziff99,Morin-Duchesne09}
where it is known as the ``cross-configuration'' probability. If a
system is critical at $p_c$, and its dual at $1-p_c$, then equation
(\ref{eq:PQ*}) holds in the scaling limit since this is merely
the statement that the cross-configuration probability is universal,
and then condition (\ref{eq:2D0D}) follows by duality. In fact, this
same argument can be made using {\it any} of the scaling limit
crossing probabilities, such as the left-right rectangular crossings
governed by Cardy's formula \cite{Cardy92,Smirnov,Lapalme01}. However, the real
power of the condition (\ref{eq:2D0D}) lies in the fact that even when
applied on small finite bases $B$, where explicit calculations are
feasible but one can expect to be nowhere near the scaling limit, it
provides very good estimates of the critical probability. Even for
bases of less than a hundred edges, we find results whose accuracy is
similar to what one obtains with state-of-the-art numerical
simulations.

\subsection{Bases and embeddings}

As mentioned above, one advantage of the redefinition (\ref{eq:2D0D})
is that we are no longer constrained to use contraction-deletion, but
may now use the transfer matrix which allows polynomials to be
calculated on much larger bases. Below we give the details of this
approach for the case of bond percolation (section~\ref{sec:tm}) and
report the results for various lattices (section~\ref{sec:bond}).

But first we discuss more carefully the bases that we have considered.
We are mainly interested in families of bases whose size can be modulated
by varying one or more integer parameters. This will in particular allow
us to study the size dependence of the resulting $p_c$.

\subsubsection{Square bases}
\label{sec:sq_bases}

\begin{figure}
\begin{center}
\includegraphics{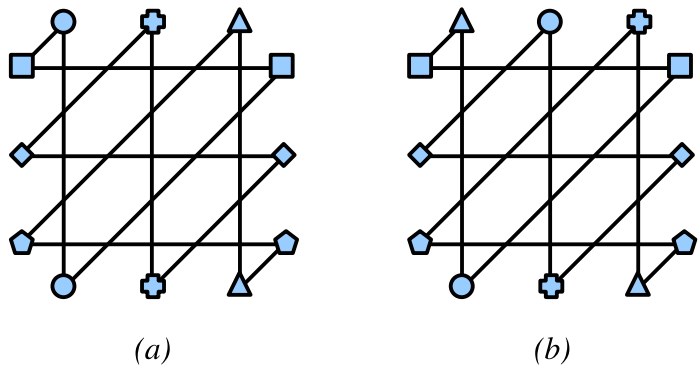}
\caption{$3 \times 3$ square bases for the kagome lattice with: a)
  straight embedding, b) a twisted embedding.}
\label{fig:squarekagome}
\end{center}
\end{figure}

An example of a square basis $B$ is shown in
Figure~\ref{fig:squarekagome}. The vertices at the tile boundaries are shared among two different
copies of $B$; we call those shared vertices the {\em terminals} of
$B$. The embedding can be visualised by pairing the terminals two by
two (shown as matching shapes in Figure~\ref{fig:squarekagome}). This
means that in the embedding a given terminal of one copy of the basis
$B_1$ is identified with the matching terminal of another copy of the
basis $B_2$. In other words, $B_1$ and $B_2$ are glued along matching
terminals. When tiling space with the basis in Figure~\ref{fig:squarekagome}a, we refer to
this as the {\em straight embedding}.

A variation of the straight embedding is to shift cyclically the
vertices along one of the sides of the square before gluing them to
those of the opposing side; we call this a {\em twisted embedding}. By
reflection symmetry, shifting cyclically $k$ steps to the right or to
the left produces identical results.  There are thus in general $1 +
\lfloor n/2 \rfloor$ inequivalent twists, corresponding to
$k=0,1,\ldots,\lfloor n/2 \rfloor$. In practice we have found that for
some---but not all---lattices the cases $(n,k) = (2,0)$ and $(n,k) =
(2,1)$ produce the same critical polynomial.

A square basis $B$ of size $n \times n$ has $n$ terminals on each of
the four sides of the square. The number of vertices and edges in $B$
are both proportional to $n^2$. In the vertex count, each terminal
counts for $1/2$ only, since it is shared among two copies of the
basis. Thus, the square basis for the kagome lattice shown in
Figure~\ref{fig:squarekagome} has $6 n^2$ edges and $3 n^2$ vertices.

One can obviously generalise this construction to rectangular bases of
size $n \times m$. For $n = m$ one recovers a square basis. For $n
\neq m$ the twists along the $n$ and $m$ directions are no longer
equivalent.

\subsubsection{Hexagonal bases}
\label{sec:hex_bases}

\begin{figure}
\begin{center}
\includegraphics{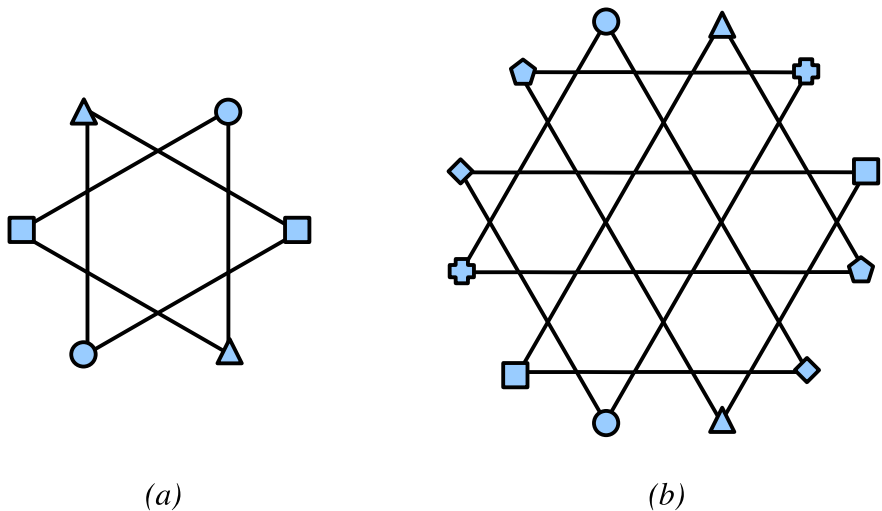}
\caption{Hexagonal bases for the kagome lattice with a) $n=1$, and b) $n=2$.}
\label{fig:hexkagome}
\end{center}
\end{figure}

When the lattice $L$ has a 3-fold rotational symmetry, one can define
as well a hexagonal embedding. Examples of this are shown in
Figure~\ref{fig:hexkagome}. Each of the six sides of the hexagon now
supports $n$ terminals. Note that it is not possible to twist the
hexagonal bases, since only the straight embedding produces a valid
tiling of two-dimensional space.

One advantage of hexagonal bases over the square bases is that they
have a lower ratio of terminals to edges, which is useful because the
number of terminals is the limiting factor in the transfer matrix
computation. For instance, for the kagome lattice one has now $6n$
terminals, $9 n^2$ vertices and $18 n^2$ edges.

Another advantage is that the hexagonal basis is designed to respect
the 3-fold rotational symmetry of the lattice.  Thus, for lattices
having this symmetry---such as the kagome and $(3,12^2)$ lattice---we
expect the hexagonal basis to yield better accuracy than the square
basis for a given number of edges. We shall come back to this point
in section~\ref{sec:bond}.

Note that one can extend this construction to generalised hexagonal
bases with $2(n_1+n_2+n_3)$ terminals, where each pair of opposing
sides of the hexagon supports $n_i$ terminals $(i=1,2,3)$. The special
case with one of the $n_i = 0$ reproduces the rectangular bases.

\section{Transfer matrix}
\label{sec:tm}

The probabilities $P(2D;B)$ and $P(0D;B)$ entering the
definition~(\ref{eq:2D0D}) of the critical polynomial can be computed
from a transfer matrix construction along the lines of
Ref.~\cite{BloteNightingale1982}.  First notice that each state of the
edges within the basis $B$ induces a set partition among the
terminals; each part (or block) in the partition consists of a subset
of terminals that are mutually connected through paths of open edges.
The key idea is to first compute the probabilities of all possible
partitions. One next groups the partitions according to their 2D, 1D
or 0D nature in order to evaluate (\ref{eq:2D0D}).

With $N$ terminals, the number of partitions respecting planarity is
given by the Catalan number
\begin{equation}
 C_N = \frac{1}{N+1} {2N \choose N} \,.
\end{equation}
For example, the $C_3 = 5$ planar partitions of the set $\{1,2,3\}$
are denoted
\begin{equation}
 (1)(2)(3) \,, \quad (12)(3) \,, \quad (13)(2) \,, \quad (1)(23) \,, \quad (123) \,,
\end{equation}
where the elements belonging to the same part are grouped inside
parentheses.

The dimension of the transfer matrix is thus $C_N$, and both time and
memory requirements are proportional to this number.%
\footnote{We assume here the use of standard sparse matrix
  factorisation techniques \cite{JacobsenCardy1998}.}
Asymptotically we have $C_N \sim 4^N$ for $N \gg 1$. Taking as an
example the kagome lattice with the $n \times n$ square basis, the
time complexity of the transfer matrix method is then $\sim 4^{4n} =
2^{8n}$. This can be compared to the contraction-deletion method,
whose number of recursive calls is $\sim 2^{6 n^2}$.

\subsection{Square bases}

\begin{figure}
\begin{center}
 \begin{tikzpicture}[scale=0.5]
  \foreach \xpos in {0,2}
  \foreach \ypos in {0,3.464}
  {
   \draw[fill,blue!15,line width=0ex] (\xpos-0.5,\ypos+0.866)--(\xpos+0.5,\ypos+2.598)--(\xpos+1.5,\ypos+0.866)--(\xpos+0.5,\ypos-0.866)--cycle;
   \draw[blue,dashed,line width=0.2ex] (\xpos-0.5,\ypos+0.866)--(\xpos+0.5,\ypos+2.598)--(\xpos+1.5,\ypos+0.866)--(\xpos+0.5,\ypos-0.866)--cycle;
   \draw[black,line width=0.3ex] (\xpos,\ypos)--(\xpos+1,\ypos)--(\xpos,\ypos+1.732)--(\xpos+1,\ypos+1.732)--cycle;
  }
  \foreach \xpos in {-1,1,3}
  \foreach \ypos in {1.732}
  {
   \draw[fill,red!15,line width=0ex] (\xpos-0.5,\ypos+0.866)--(\xpos+0.5,\ypos+2.598)--(\xpos+1.5,\ypos+0.866)--(\xpos+0.5,\ypos-0.866)--cycle;
   \draw[blue,dashed,line width=0.2ex] (\xpos-0.5,\ypos+0.866)--(\xpos+0.5,\ypos+2.598)--(\xpos+1.5,\ypos+0.866)--(\xpos+0.5,\ypos-0.866)--cycle;
   \draw[black,line width=0.3ex] (\xpos,\ypos)--(\xpos+1,\ypos)--(\xpos,\ypos+1.732)--(\xpos+1,\ypos+1.732)--cycle;
  }
  \foreach \xpos in {1}
  \foreach \ypos in {-1.732,5.196}
  {
   \draw[fill,red!15,line width=0ex] (\xpos-0.5,\ypos+0.866)--(\xpos+0.5,\ypos+2.598)--(\xpos+1.5,\ypos+0.866)--(\xpos+0.5,\ypos-0.866)--cycle;
   \draw[blue,dashed,line width=0.2ex] (\xpos-0.5,\ypos+0.866)--(\xpos+0.5,\ypos+2.598)--(\xpos+1.5,\ypos+0.866)--(\xpos+0.5,\ypos-0.866)--cycle;
   \draw[black,line width=0.3ex] (\xpos,\ypos)--(\xpos+1,\ypos)--(\xpos,\ypos+1.732)--(\xpos+1,\ypos+1.732)--cycle;
  }
 \draw (-1,1.732) node[below left] {$1$};
 \draw (0,0) node[below left] {$2$};
 \draw (1,-1.732) node[below left] {$3$};
 \draw (2,-1.732) node[below right] {$4$};
 \draw (3,0) node[below right] {$5$};
 \draw (4,1.732) node[below right] {$6$};
 \draw (-1,3.464) node[above left] {$1'$};
 \draw (0,5.196) node[above left] {$2'$};
 \draw (1,6.928) node[above left] {$3'$};
 \draw (2,6.928) node[above right] {$4'$};
 \draw (3,5.196) node[above right] {$5'$};
 \draw (4,3.464) node[above right] {$6'$};
 \begin{scope}[xshift=12cm,yshift=1.732cm]
  \draw (-0.5,0.866) node[left] {${\sf B}_i =$};
  \draw (0,0) node[below left] {$i$};
  \draw (1,0) node[below right] {$i+1$};
  \draw (0,1.732) node[above left] {$i'$};
  \draw (1,1.732) node[above right] {$i'+1$};
  \foreach \xpos in {0}
  \foreach \ypos in {0}
  {
   \draw[fill,red!0,line width=0ex] (\xpos-0.5,\ypos+0.866)--(\xpos+0.5,\ypos+2.598)--(\xpos+1.5,\ypos+0.866)--(\xpos+0.5,\ypos-0.866)--cycle;
   \draw[blue,dashed,line width=0.2ex] (\xpos-0.5,\ypos+0.866)--(\xpos+0.5,\ypos+2.598)--(\xpos+1.5,\ypos+0.866)--(\xpos+0.5,\ypos-0.866)--cycle;
   \draw[black,line width=0.3ex] (\xpos,\ypos)--(\xpos+1,\ypos)--(\xpos,\ypos+1.732)--(\xpos+1,\ypos+1.732)--cycle;
  }
 \end{scope}
 \end{tikzpicture}
\end{center}
 \caption{Transfer matrix construction for the kagome lattice on
   an $n \times n$ square basis, here with $n=3$. The operator ${\sf
     B}_i$ adds six edges to the lattice.}
 \label{fig:kagomeTM}
\end{figure}
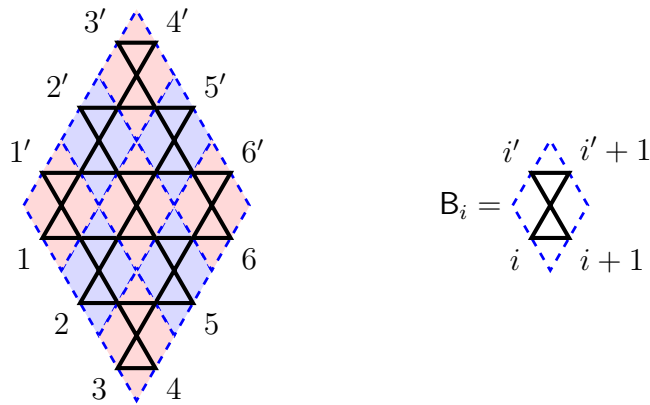

Our transfer matrix construction is most easily explained on a specific
example. So consider the kagome lattice with the $n \times n$ square basis;
the case $n=3$ is shown in Figure~\ref{fig:kagomeTM}.

The transfer matrix ${\sf T}$ constructs the lattice from the bottom
to the top, while keeping track of the Boltzmann weight of each
partition of the terminals.  The bottom terminals are denoted
$1,2,\ldots,2n$ and the top terminals $1',2',\ldots,2n'$. At the
beginning of the process the top and bottom are identified, so the
initial state $|{\rm i} \rangle$ on which ${\sf T}$ acts is the partition
$(1\,1')(2\,2')\cdots(2n\,2n')$ with weight $1$.

We now define two kinds of operators acting on a partition
\cite{SalasSokal2001}:
\begin{itemize}
 \item The join operator ${\sf J}_i$ amalgamates the parts to which
   the top terminals $i'$ and $i'+1$ belong. In particular, on
   partitions in which those two terminals already belong to the same part,
   ${\sf J}_i$ acts as the identity operator. Note that if some parts
   contain both bottom and top terminals, the action of ${\sf J}_i$
   can also affect the connections among the bottom terminals.
 \item The detach operator ${\sf D}_i$ detaches the top terminal $i'$
   from its part and transforms it into a singleton in the
   partition. In particular, if that terminal was already a singleton,
   ${\sf D}_i$ acts as the identity operator.
\end{itemize}
{}From these two basic operators and the identity operator ${\sf I}$
we now define an operator
\begin{equation}
 {\sf H}_i = {\sf I} + v {\sf J}_i
\end{equation}
that adds a horizontal edge to the lattice. The word ``horizontal''
refers to a drawing of the lattice where the top terminals $i'$ and
$i'+1$ are horizontally aligned; otherwise the edge would be
better described as ``diagonal''.  Note that ${\sf H}_i$ attaches a
weight $1$ (resp.\ $v$) to a closed (resp.\ open) horizontal edge,
as required. Similarly we define
\begin{equation}
 {\sf V}_i = v {\sf I} + {\sf D}_i
\end{equation}
that adds a vertical edge between $i'$ and $i''$, where $i'$
(resp.\ $i''$) denotes the corresponding top terminal before
(resp.\ after) the action of ${\sf V}_i$. To simplify the notation, it
is convenient to assume that following the action of ${\sf V}_i$ we
relabel $i''$ as $i'$. The word ``vertical'' refers to a drawing of the
lattice where $i'$ and $i''$ are vertically aligned.

The fundamental building block of the lattice shown on the right of
Figure~\ref{fig:kagomeTM} is then constructed by the composite
operator
\begin{equation}
 {\sf B}_i = {\sf H}_i {\sf V}_i {\sf H}_i {\sf D}_{i+1} {\sf H}_i {\sf V}_i {\sf H}_i \,,
 \qquad \mbox{Kagome lattice.}
\end{equation}
The whole lattice $B$ is finally obtained by adding successive rows
(for clarity shown in alternating hues on the left of
Figure~\ref{fig:kagomeTM}) of ${\sf B}_i$. The transfer matrix then reads
\begin{equation}
 {\sf T} = \prod_{y=1}^{n-1} \prod_{x=1}^{y} {\sf B}_{n-y-1+2x} \times
           \prod_{y=1}^n \prod_{x=0}^{n-y} {\sf B}_{y+2x}
\end{equation}
and the final state
\begin{equation}
 |{\rm f}\rangle = {\sf T} |{\rm i}\rangle
 \label{final_state}
\end{equation}
contains all possible partitions among the $4n$ terminals along with their
respective Boltzmann weights.

\subsubsection{Other lattices}

The extension to the other lattices considered in this paper is very
simple: it suffices to change the definition of the operator ${\sf
  B}_i$, while leaving the remainder of the construction unchanged.%
\footnote{In practice, when implementing this algorithm on a computer,
  this implies that only a few lines of code have to be modified to
  change the lattice.}

\begin{figure}
\begin{center}
 \begin{tikzpicture}[scale=0.5]
  \foreach \xpos in {0,3.414}
  \foreach \ypos in {0,3.414}
  {
   \draw[fill,blue!15,line width=0ex] (\xpos-0.5,\ypos+0.5)--(\xpos+1.207,\ypos+2.207)--(\xpos+2.914,\ypos+0.5)--(\xpos+1.207,\ypos-1.207)--cycle;
   \draw[blue,dashed,line width=0.2ex] (\xpos-0.5,\ypos+0.5)--(\xpos+1.207,\ypos+2.207)--(\xpos+2.914,\ypos+0.5)--(\xpos+1.207,\ypos-1.207)--cycle;
   \draw[black,line width=0.3ex] (\xpos,\ypos)--(\xpos+0.707,\ypos+0.707)--(\xpos+1.707,\ypos+0.707)--(\xpos+2.414,\ypos);
   \draw[black,line width=0.3ex] (\xpos+0.707,\ypos+0.707)--(\xpos+0.707,\ypos+1.707)--(\xpos+1.707,\ypos+1.707)--(\xpos+1.707,\ypos+0.707);
  }
  \foreach \xpos in {-1.707,1.707,5.121}
  \foreach \ypos in {1.707}
  {
   \draw[fill,red!15,line width=0ex] (\xpos-0.5,\ypos+0.5)--(\xpos+1.207,\ypos+2.207)--(\xpos+2.914,\ypos+0.5)--(\xpos+1.207,\ypos-1.207)--cycle;
   \draw[blue,dashed,line width=0.2ex] (\xpos-0.5,\ypos+0.5)--(\xpos+1.207,\ypos+2.207)--(\xpos+2.914,\ypos+0.5)--(\xpos+1.207,\ypos-1.207)--cycle;
   \draw[black,line width=0.3ex] (\xpos,\ypos)--(\xpos+0.707,\ypos+0.707)--(\xpos+1.707,\ypos+0.707)--(\xpos+2.414,\ypos);
   \draw[black,line width=0.3ex] (\xpos+0.707,\ypos+0.707)--(\xpos+0.707,\ypos+1.707)--(\xpos+1.707,\ypos+1.707)--(\xpos+1.707,\ypos+0.707);
  }
  \foreach \xpos in {1.707}
  \foreach \ypos in {-1.707,5.121}
  {
   \draw[fill,red!15,line width=0ex] (\xpos-0.5,\ypos+0.5)--(\xpos+1.207,\ypos+2.207)--(\xpos+2.914,\ypos+0.5)--(\xpos+1.207,\ypos-1.207)--cycle;
   \draw[blue,dashed,line width=0.2ex] (\xpos-0.5,\ypos+0.5)--(\xpos+1.207,\ypos+2.207)--(\xpos+2.914,\ypos+0.5)--(\xpos+1.207,\ypos-1.207)--cycle;
   \draw[black,line width=0.3ex] (\xpos,\ypos)--(\xpos+0.707,\ypos+0.707)--(\xpos+1.707,\ypos+0.707)--(\xpos+2.414,\ypos);
   \draw[black,line width=0.3ex] (\xpos+0.707,\ypos+0.707)--(\xpos+0.707,\ypos+1.707)--(\xpos+1.707,\ypos+1.707)--(\xpos+1.707,\ypos+0.707);
  }
 \draw (-1.707,1.707) node[below left] {$1$};
 \draw (0,0) node[below left] {$2$};
 \draw (1.707,-1.707) node[below left] {$3$};
 \draw (4.121,-1.707) node[below right] {$4$};
 \draw (5.828,0) node[below right] {$5$};
 \draw (7.535,1.707) node[below right] {$6$};
 \draw (-1.000,3.414) node[above left] {$1'$};
 \draw (0.707,5.121) node[above left] {$2'$};
 \draw (2.414,6.828) node[above left] {$3'$};
 \draw (3.414,6.828) node[above right] {$4'$};
 \draw (5.121,5.121) node[above right] {$5'$};
 \draw (6.828,3.414) node[above right] {$6'$};
  \begin{scope}[xshift=15.5cm,yshift=1.707cm]
  \draw (-0.5,0.5) node[left] {${\sf B}_i =$};
  \draw (0,0) node[below left] {$i$};
  \draw (2.414,0) node[below right] {$i+1$};
  \draw (0.707,1.707) node[above left] {$i'$};
  \draw (1.707,1.707) node[above right] {$i'+1$};
  \foreach \xpos in {0}
  \foreach \ypos in {0}
  {
   \draw[fill,red!0,line width=0ex] (\xpos-0.5,\ypos+0.5)--(\xpos+1.207,\ypos+2.207)--(\xpos+2.914,\ypos+0.5)--(\xpos+1.207,\ypos-1.207)--cycle;
   \draw[blue,dashed,line width=0.2ex] (\xpos-0.5,\ypos+0.5)--(\xpos+1.207,\ypos+2.207)--(\xpos+2.914,\ypos+0.5)--(\xpos+1.207,\ypos-1.207)--cycle;
   \draw[black,line width=0.3ex] (\xpos,\ypos)--(\xpos+0.707,\ypos+0.707)--(\xpos+1.707,\ypos+0.707)--(\xpos+2.414,\ypos);
   \draw[black,line width=0.3ex] (\xpos+0.707,\ypos+0.707)--(\xpos+0.707,\ypos+1.707)--(\xpos+1.707,\ypos+1.707)--(\xpos+1.707,\ypos+0.707);
  }
 \end{scope}
 \end{tikzpicture}
\end{center}
 \caption{Transfer matrix construction for the $(4,8^2)$ lattice on
   an $n \times n$ square basis, here with $n=3$. The operator ${\sf
     B}_i$ adds six edges to the lattice.}
 \label{fig:foureightTM}
\end{figure}
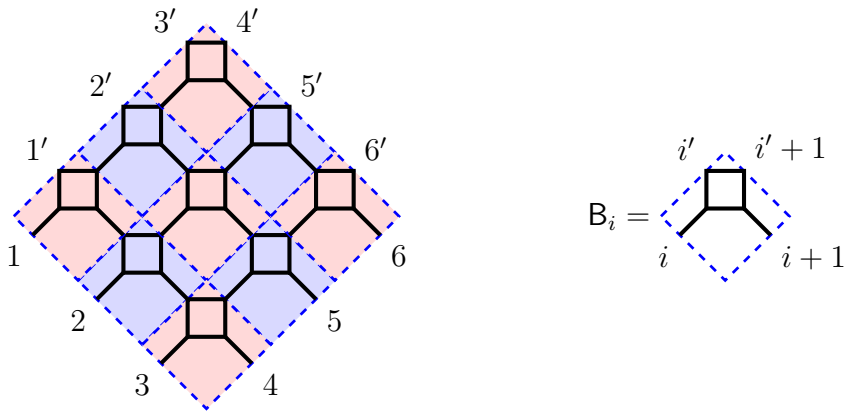

The square basis for the $(4,8^2)$ lattice is shown in Figure~\ref{fig:foureightTM}.
Its fundamental building block now has the expression
\begin{equation}
 {\sf B}_i = {\sf H}_i {\sf V}_i {\sf V}_{i+1} {\sf H}_i {\sf V}_i {\sf V}_{i+1} \,,
 \qquad \mbox{$(4,8^2)$ lattice.}
\end{equation}

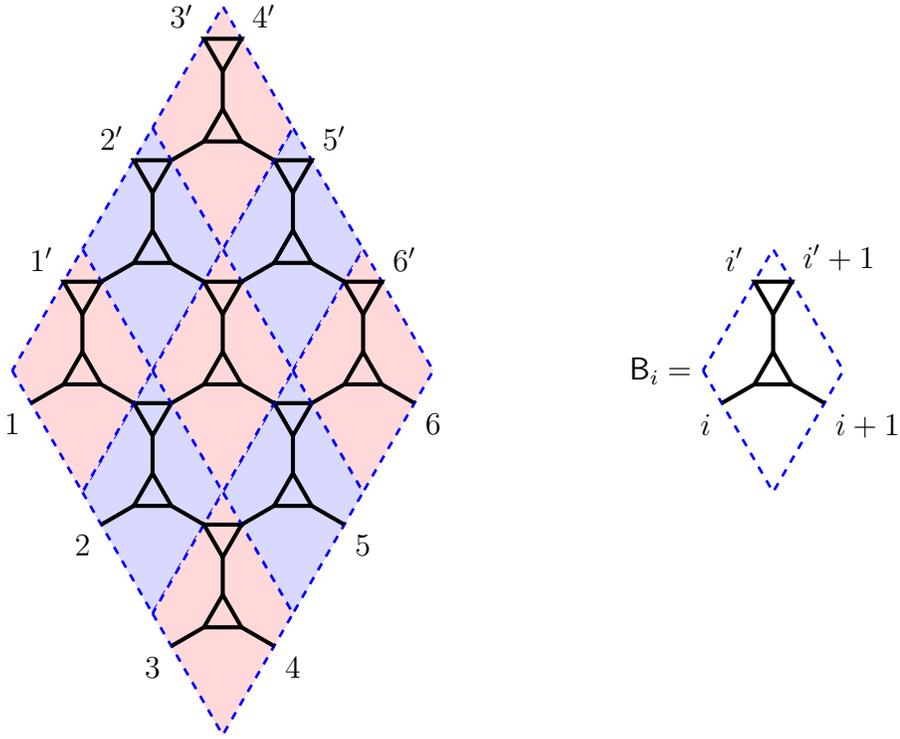
\begin{figure}
\begin{center}
 \begin{tikzpicture}[scale=0.5]
  \foreach \xpos in {0,3.732}
  \foreach \ypos in {0,6.464}
  {
   \draw[fill,blue!15,line width=0ex] (\xpos-0.5,\ypos+0.866)--(\xpos+1.366,\ypos-2.366)--(\xpos+3.232,\ypos+0.866)--(\xpos+1.366,\ypos+4.098)--cycle;
   \draw[blue,dashed,line width=0.2ex] (\xpos-0.5,\ypos+0.866)--(\xpos+1.366,\ypos-2.366)--(\xpos+3.232,\ypos+0.866)--(\xpos+1.366,\ypos+4.098)--cycle;
   \draw[black,line width=0.3ex] (\xpos,\ypos)--(\xpos+0.866,\ypos+0.5)--(\xpos+1.866,\ypos+0.5)--(\xpos+1.366,\ypos+1.366)--(\xpos+0.866,\ypos+0.5);
   \draw[black,line width=0.3ex] (\xpos+1.866,\ypos+0.5)--(\xpos+2.732,\ypos);
   \draw[black,line width=0.3ex] (\xpos+1.366,\ypos+1.366)--(\xpos+1.366,\ypos+2.366)--(\xpos+1.866,\ypos+3.232)--(\xpos+0.866,\ypos+3.232)--(\xpos+1.366,\ypos+2.366);
  }
  \foreach \xpos in {-1.866,1.866,5.598}
  \foreach \ypos in {3.232}
  {
   \draw[fill,red!15,line width=0ex] (\xpos-0.5,\ypos+0.866)--(\xpos+1.366,\ypos-2.366)--(\xpos+3.232,\ypos+0.866)--(\xpos+1.366,\ypos+4.098)--cycle;
   \draw[blue,dashed,line width=0.2ex] (\xpos-0.5,\ypos+0.866)--(\xpos+1.366,\ypos-2.366)--(\xpos+3.232,\ypos+0.866)--(\xpos+1.366,\ypos+4.098)--cycle;
   \draw[black,line width=0.3ex] (\xpos,\ypos)--(\xpos+0.866,\ypos+0.5)--(\xpos+1.866,\ypos+0.5)--(\xpos+1.366,\ypos+1.366)--(\xpos+0.866,\ypos+0.5);
   \draw[black,line width=0.3ex] (\xpos+1.866,\ypos+0.5)--(\xpos+2.732,\ypos);
   \draw[black,line width=0.3ex] (\xpos+1.366,\ypos+1.366)--(\xpos+1.366,\ypos+2.366)--(\xpos+1.866,\ypos+3.232)--(\xpos+0.866,\ypos+3.232)--(\xpos+1.366,\ypos+2.366);
  }
  \foreach \xpos in {1.866}
  \foreach \ypos in {-3.232,9.696}
  {
   \draw[fill,red!15,line width=0ex] (\xpos-0.5,\ypos+0.866)--(\xpos+1.366,\ypos-2.366)--(\xpos+3.232,\ypos+0.866)--(\xpos+1.366,\ypos+4.098)--cycle;
   \draw[blue,dashed,line width=0.2ex] (\xpos-0.5,\ypos+0.866)--(\xpos+1.366,\ypos-2.366)--(\xpos+3.232,\ypos+0.866)--(\xpos+1.366,\ypos+4.098)--cycle;
   \draw[black,line width=0.3ex] (\xpos,\ypos)--(\xpos+0.866,\ypos+0.5)--(\xpos+1.866,\ypos+0.5)--(\xpos+1.366,\ypos+1.366)--(\xpos+0.866,\ypos+0.5);
   \draw[black,line width=0.3ex] (\xpos+1.866,\ypos+0.5)--(\xpos+2.732,\ypos);
   \draw[black,line width=0.3ex] (\xpos+1.366,\ypos+1.366)--(\xpos+1.366,\ypos+2.366)--(\xpos+1.866,\ypos+3.232)--(\xpos+0.866,\ypos+3.232)--(\xpos+1.366,\ypos+2.366);
  }
 \draw (-1.866,3.232) node[below left] {$1$};
 \draw (0,0) node[below left] {$2$};
 \draw (1.866,-3.232) node[below left] {$3$};
 \draw (4.598,-3.232) node[below right] {$4$};
 \draw (6.464,0) node[below right] {$5$};
 \draw (8.330,3.232) node[below right] {$6$};
 \draw (-1.000,6.464) node[above left] {$1'$};
 \draw (0.866,9.696) node[above left] {$2'$};
 \draw (2.732,12.928) node[above left] {$3'$};
 \draw (3.732,12.928) node[above right] {$4'$};
 \draw (5.598,9.696) node[above right] {$5'$};
 \draw (7.464,6.464) node[above right] {$6'$};
  \begin{scope}[xshift=16.5cm,yshift=3.232cm]
  \draw (-0.5,0.866) node[left] {${\sf B}_i =$};
  \draw (0,0) node[below left] {$i$};
  \draw (2.732,0) node[below right] {$i+1$};
  \draw (0.866,3.232) node[above left] {$i'$};
  \draw (1.866,3.232) node[above right] {$i'+1$};
  \foreach \xpos in {0}
  \foreach \ypos in {0}
  {
   \draw[fill,red!0,line width=0ex] (\xpos-0.5,\ypos+0.866)--(\xpos+1.366,\ypos-2.366)--(\xpos+3.232,\ypos+0.866)--(\xpos+1.366,\ypos+4.098)--cycle;
   \draw[blue,dashed,line width=0.2ex] (\xpos-0.5,\ypos+0.866)--(\xpos+1.366,\ypos-2.366)--(\xpos+3.232,\ypos+0.866)--(\xpos+1.366,\ypos+4.098)--cycle;
   \draw[black,line width=0.3ex] (\xpos,\ypos)--(\xpos+0.866,\ypos+0.5)--(\xpos+1.866,\ypos+0.5)--(\xpos+1.366,\ypos+1.366)--(\xpos+0.866,\ypos+0.5);
   \draw[black,line width=0.3ex] (\xpos+1.866,\ypos+0.5)--(\xpos+2.732,\ypos);
   \draw[black,line width=0.3ex] (\xpos+1.366,\ypos+1.366)--(\xpos+1.366,\ypos+2.366)--(\xpos+1.866,\ypos+3.232)--(\xpos+0.866,\ypos+3.232)--(\xpos+1.366,\ypos+2.366);
  }
 \end{scope}
 \end{tikzpicture}
\end{center}
 \caption{Transfer matrix construction for the $(3,12^2)$ lattice on
   an $n \times n$ square basis, here with $n=3$. The operator ${\sf
     B}_i$ adds nine edges to the lattice.}
 \label{fig:threetwelveTM}
\end{figure}

As a last example, consider the $(3,12^2)$ lattice with the square basis depicted
in Figure~\ref{fig:threetwelveTM}. We find in this case
\begin{equation}
 {\sf B}_i = {\sf H}_i {\sf V}_i {\sf H}_i {\sf V}_i {\sf D}_{i+1}
             {\sf H}_i {\sf V}_i {\sf H}_i {\sf V}_i {\sf V}_{i+1} \,,
 \qquad \mbox{$(3,12^2)$ lattice.}
\end{equation}

\subsection{Hexagonal bases}

Because of their 3-fold rotational symmetry, it is also interesting to study the
kagome and $(3,12^2)$ lattice with a hexagonal basis. We now describe how to
adapt the transfer matrix construction to this case.

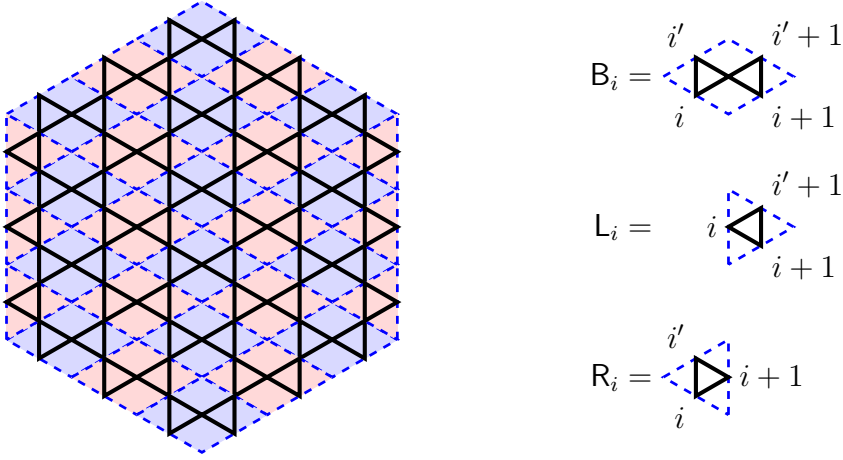
\begin{figure}
\begin{center}
 \begin{tikzpicture}[scale=0.5]
  \foreach \xpos in {-3.464,0,3.464}
  \foreach \ypos in {2,4,6,8}
  {
   \draw[fill,blue!15,line width=0ex] (\xpos,\ypos)--(\xpos+1.732,\ypos+1)--(\xpos+3.464,\ypos)--(\xpos+1.732,\ypos-1)--cycle;
   \draw[blue,dashed,line width=0.2ex] (\xpos,\ypos)--(\xpos+1.732,\ypos+1)--(\xpos+3.464,\ypos)--(\xpos+1.732,\ypos-1)--cycle;
   \draw[black,line width=0.3ex] (\xpos+0.866,\ypos-0.5)--(\xpos+2.598,\ypos+0.5)--(\xpos+2.598,\ypos-0.5)--(\xpos+0.866,\ypos+0.5)--cycle;
  }
  \foreach \xpos in {-1.732,1.732}
  \foreach \ypos in {1,3,5,7,9}
  {
   \draw[fill,red!15,line width=0ex] (\xpos,\ypos)--(\xpos+1.732,\ypos+1)--(\xpos+3.464,\ypos)--(\xpos+1.732,\ypos-1)--cycle;
   \draw[blue,dashed,line width=0.2ex] (\xpos,\ypos)--(\xpos+1.732,\ypos+1)--(\xpos+3.464,\ypos)--(\xpos+1.732,\ypos-1)--cycle;
   \draw[black,line width=0.3ex] (\xpos+0.866,\ypos-0.5)--(\xpos+2.598,\ypos+0.5)--(\xpos+2.598,\ypos-0.5)--(\xpos+0.866,\ypos+0.5)--cycle;
  }
  \foreach \xpos in {0}
  \foreach \ypos in {0,10}
  {
   \draw[fill,blue!15,line width=0ex] (\xpos,\ypos)--(\xpos+1.732,\ypos+1)--(\xpos+3.464,\ypos)--(\xpos+1.732,\ypos-1)--cycle;
   \draw[blue,dashed,line width=0.2ex] (\xpos,\ypos)--(\xpos+1.732,\ypos+1)--(\xpos+3.464,\ypos)--(\xpos+1.732,\ypos-1)--cycle;
   \draw[black,line width=0.3ex] (\xpos+0.866,\ypos-0.5)--(\xpos+2.598,\ypos+0.5)--(\xpos+2.598,\ypos-0.5)--(\xpos+0.866,\ypos+0.5)--cycle;
  }
  \foreach \xpos in {-3.464}
  \foreach \ypos in {3,5,7}
  {
   \draw[fill,red!15,line width=0ex] (\xpos,\ypos-1)--(\xpos,\ypos+1)--(\xpos+1.732,\ypos)--cycle;
   \draw[blue,dashed,line width=0.2ex] (\xpos,\ypos-1)--(\xpos,\ypos+1)--(\xpos+1.732,\ypos)--cycle;
   \draw[black,line width=0.3ex] (\xpos,\ypos)--(\xpos+0.866,\ypos+0.5)--(\xpos+0.866,\ypos-0.5)--cycle;
  }
  \foreach \xpos in {6.928}
  \foreach \ypos in {3,5,7}
  {
   \draw[fill,red!15,line width=0ex] (\xpos,\ypos-1)--(\xpos,\ypos+1)--(\xpos-1.732,\ypos)--cycle;
   \draw[blue,dashed,line width=0.2ex] (\xpos,\ypos-1)--(\xpos,\ypos+1)--(\xpos-1.732,\ypos)--cycle;
   \draw[black,line width=0.3ex] (\xpos,\ypos)--(\xpos-0.866,\ypos+0.5)--(\xpos-0.866,\ypos-0.5)--cycle;
  }
 \begin{scope}[xshift=14cm,yshift=9cm]
  \draw (0,0) node[left] {${\sf B}_i =$};
  \draw (0.866,-0.5) node[below left] {$i$};
  \draw (2.598,-0.5) node[below right] {$i+1$};
  \draw (0.866,0.5) node[above left] {$i'$};
  \draw (2.598,0.5) node[above right] {$i'+1$};
  \foreach \xpos in {0}
  \foreach \ypos in {0}
  {
   \draw[fill,blue!0,line width=0ex] (\xpos,\ypos)--(\xpos+1.732,\ypos+1)--(\xpos+3.464,\ypos)--(\xpos+1.732,\ypos-1)--cycle;
   \draw[blue,dashed,line width=0.2ex] (\xpos,\ypos)--(\xpos+1.732,\ypos+1)--(\xpos+3.464,\ypos)--(\xpos+1.732,\ypos-1)--cycle;
   \draw[black,line width=0.3ex] (\xpos+0.866,\ypos-0.5)--(\xpos+2.598,\ypos+0.5)--(\xpos+2.598,\ypos-0.5)--(\xpos+0.866,\ypos+0.5)--cycle;
  }
 \end{scope}
 \begin{scope}[xshift=14cm,yshift=5cm]
  \draw (0,0) node[left] {${\sf L}_i =$};
  \draw (1.732,0) node[left] {$i$};
  \draw (2.598,-0.5) node[below right] {$i+1$};
  \draw (2.598,0.5) node[above right] {$i'+1$};
  \foreach \xpos in {1.732}
  \foreach \ypos in {0}
  {
   \draw[fill,red!0,line width=0ex] (\xpos,\ypos-1)--(\xpos,\ypos+1)--(\xpos+1.732,\ypos)--cycle;
   \draw[blue,dashed,line width=0.2ex] (\xpos,\ypos-1)--(\xpos,\ypos+1)--(\xpos+1.732,\ypos)--cycle;
   \draw[black,line width=0.3ex] (\xpos,\ypos)--(\xpos+0.866,\ypos+0.5)--(\xpos+0.866,\ypos-0.5)--cycle;
  }
 \end{scope}
 \begin{scope}[xshift=14cm,yshift=1cm]
  \draw (0,0) node[left] {${\sf R}_i =$};
  \draw (1.732,0) node[right] {$i+1$};
  \draw (0.866,-0.5) node[below left] {$i$};
  \draw (0.866,0.5) node[above left] {$i'$};
  \foreach \xpos in {1.732}
  \foreach \ypos in {0}
  {
   \draw[fill,red!0,line width=0ex] (\xpos,\ypos-1)--(\xpos,\ypos+1)--(\xpos-1.732,\ypos)--cycle;
   \draw[blue,dashed,line width=0.2ex] (\xpos,\ypos-1)--(\xpos,\ypos+1)--(\xpos-1.732,\ypos)--cycle;
   \draw[black,line width=0.3ex] (\xpos,\ypos)--(\xpos-0.866,\ypos+0.5)--(\xpos-0.866,\ypos-0.5)--cycle;
  }
 \end{scope}
 \end{tikzpicture}
\end{center}
\caption{Kagome lattice on a hexagonal basis of size $n$, here with $n=3$. The operator
${\sf B}_i$ adds six edges to the lattice, while the left and right boundary operators,
${\sf L}_i$ and ${\sf R}_i$, each add three.}
\label{fig:kagome_hex}
\end{figure}

Consider as an example the kagome lattice with the hexagonal basis of
size $n$; the case $n=3$ is shown in
Figure~\ref{fig:kagome_hex}. There are now $6n$ terminals.  Those on
the two bottom sides (resp.\ the two top sides) of the hexagon are
labelled $1,2,\ldots,2n$ (resp.\ $1',2',\ldots,2n'$), just as in the
case of the square basis. We describe below how the remaining
terminals on the left and right sides of the hexagon are to be
handled. The transfer matrix ${\sf T}$ still constructs the lattice
from the bottom to the top.

The expression for the building block ${\sf B}_i$ now needs some
modification, since the orientation of the bow tie motif with respect
to the transfer direction (invariably upwards) has been changed. One
easy option would be to handle the centre of the bow tie as an extra
point---we would then label the three points $i$, $i+1$ and
$i+2$)---and use the expression ${\bf B}_i = {\sf D}_{i+1} {\sf
  H}_{i+1} {\sf H}_i {\sf V}_{i+2} {\sf V}_i {\sf H}_{i+1} {\sf H}_i$.
It is however more efficient to avoid introducing the centre point
into the partition (and keep the usual labelling $i$, $i+1$ as shown
on the right of Figure~\ref{fig:kagome_hex}). The expression for ${\sf
  B}_i$ can then be found by computing the final state
(\ref{final_state}) for the $1 \times 1$ square basis
and rotating the labels (we denote here $j = i+1$):
\begin{eqnarray}
 {\sf B}_i &=& (v^6 + 6 v^5 + 9 v^4) (iji'j') + (2v^4+6v^3+v^2) (ii')(jj') \nonumber \\
           &+& (v^4 + 3 v^3) \big[ (i)(ji'j') +
               (j)(ii'j') + (i')(ijj') + (j')(iji') \big] \nonumber \\
           &+& (v^3 + 5 v^2 + v) \big[ (ii')(j)(j') + (i)(i')(jj') \big] +
               (4v+1) (i)(j)(i')(j') \nonumber \\
           &+& v^2 \left[ (i)(j)(i'j') + (ij)(i')(j') + (i)(i'j)(j') + (i')(ij')(j) \right] \,.
 \label{newBkagome}
\end{eqnarray}
where a bracketed operator, for example $(ii')(jj')$, creates a bow-tie between $i$ and $j$ with the indicated partition of its four bounding vertices.
On the boundary of the hexagon we need the further operators
\begin{eqnarray}
 {\sf L}_i &=& {\sf H}_i {\sf V}_{i+1} {\sf H}_i \,, \\
 {\sf R}_i &=& {\sf H}_i {\sf V}_i {\sf H}_i \,.
\end{eqnarray}
The transfer matrix that builds the whole hexagon then reads
\begin{eqnarray}
 {\sf T} &=& \prod_{y=1}^{n-1} \prod_{x=1}^{y} {\sf B}_{n-y-1+2x} \nonumber \\
    &\times& \prod_{y=1}^n \left( \prod_{x=1}^n {\sf B}_{2x-1} \times
             {\sf L}_0 \prod_{x=1}^{n-1} {\sf B}_{2x} \times R_{2n} \right) \times
           \prod_{y=1}^n \prod_{x=0}^{n-y} {\sf B}_{y+2x} \,.
 \label{hexT}
\end{eqnarray}

Regarding the handling of the boundary points, a small remark is in
order. In (\ref{hexT}) these have been denoted simply $0$ (on the
left) and $2n+1$ (on the right). In the initial state $|{\rm
  i}\rangle$, both $0$ and $2n+1$ are singletons. After each factor in
the middle product over $y$ the two boundary labels have to be stored,
so that in the final state (\ref{final_state}) the partitions indeed
involve all $6n$ terminals. To avoid introducing a cumbersome
notation, we understand implicitly that this storing is performed when
expanding the product (\ref{hexT}).

\subsubsection{Other lattices} \label{sec:OL}

The $(3,12^2)$ lattice can be handled similarly by rotating ${\sf
  B}_i$ shown in the right part of Figure~\ref{fig:threetwelveTM}
through angle $\pi/2$ clockwise. The left (resp.\ right) boundary
operator ${\sf L}_i$ (resp.\ ${\sf R}_i$) then consists of the four
rightmost (resp.\ five leftmost) edges in the rotated ${\sf B}_i$.

Explicitly we find
\begin{eqnarray}
 {\sf B}_i &=& (v^9 + 6 v^8 + 9 v^7) (iji'j') +
   (v^7 + 3 v^6) \left[ (iji')(j') + (iji')(j) \right] \nonumber \\
 &+& (v^8 + 7 v^7 + 13 v^6 + 3 v^5) \left[ (i)(ji'j') + (i')(ijj') \right] \nonumber \\
 &+& (v^8 + 8 v^7 + 17 v^6 + 7 v^5 + v^4) (ii')(jj') \nonumber \\
 &+& (3 v^6 + 12 v^5 + 6 v^4 + v^3) (ii')(j)(j') \\
 &+& (3 v^7 + 28 v^6 + 77 v^5 + 70 v^4 + 34 v^3 + 9 v^2 + v) (i)(i')(jj') \nonumber \\
 &+& (v^6 + 4 v^5 + v^4)
     \left[ (i)(j)(i'j') + (ij)(i')(j') + (i)(i'j)(j') + (i')(ij')(j) \right] \nonumber \\
 &+& (8 v^5 + 45 v^4 + 49 v^3 + 27 v^2 + 8 v + 1) (i)(j)(i')(j') \nonumber
\end{eqnarray}
along with
\begin{eqnarray}
 {\sf L}_i &=& (v^4 + 3 v^3) (ijj') + (v^3 + 4 v^2 + v) (i)(jj') \nonumber \\
 &+& v^2 \left[ (ij)(j') + (ij')(j) \right] + (3 v + 1) (i)(j)(j')
\end{eqnarray}
and
\begin{eqnarray}
 {\sf R}_i &=& (v^5 + 3 v^4) (ii'j) + (v^4 + 4 v^3 + v^2) \left[ (ij)(i') + (i)(i'j) \right]
               \nonumber \\
 &+& v^3 (ii')(j) + (v^3 + 8 v^2 + 5 v + 1) (i)(j)(i') \,.
\end{eqnarray}

The other problem we can handle with this construction is site percolation on the hexagonal lattice. Here, a ``bow-tie'' consists only of two sites, which replace the triangles of the kagome lattice. Now many of the weights in the operator ${\sf B}_i$ are zero, as those partitions are not possible, and the remaining terms are fairly simple:
\begin{equation}
 {\sf B}_i = v^2 (iji'j') + v \left[ (ii')(j)(j') + (i)(i')(jj') \right] + 1 (i)(j)(i')(j') \label{eq:hexB}
\end{equation}
with
\begin{equation}
  {\sf L}_i = v (ijj')+1 (i)(j)(j')
\end{equation}
and
\begin{equation}
{\sf R}_i = v (ii'j)+1 (i)(i')(j') \ . \label{eq:hexR}
\end{equation}
\subsection{Distinguishing 2D, 1D and 0D partitions}

We now explain how each partition entering the final state
(\ref{final_state}) can be assigned the correct homotopy (0D, 1D or
2D) in order to make possible the application of the main result
(\ref{eq:2D0D}). The definition of homotopy that we have given in
section~\ref{sec:altdef} is not very practical, because it refers to
the connectivity properties between two arbitrarily separated copies
of the basis, $B_1$ and $B_2$. The purpose of this section is to
provide an operational determination of the homotopy using just intrinsic
properties of $B$.

Each partition of the set of $N$ terminals can be represented as a
planar hypergraph on $N$ vertices, where each part of size $k>1$ in
the partition corresponds to a hyperedge of degree $d = k-1$ in the
hypergraph. Because of the planarity we can obtain yet another
representation as an ordinary graph on $2N$ vertices with precisely
$N$ ordinary ($d=1$) edges. We now detail this construction, which is
completely analogous to a well-known \cite{BaxterKellandWu1976}
equivalence for the partition function of the Potts model defined on a
planar graph $G$ that can be represented, on the one hand, in terms of
Fortuin-Kasteleyn clusters \cite{FK1972} on $G$ and, on the other
hand, as a loop model on the medial graph ${\cal M}(G)$.

The hypergraph can be drawn inside the frame (the outer boundary of
the shaded areas in Figures~\ref{fig:kagomeTM} and
\ref{fig:kagome_hex}) on which the $N$ terminals live. Here we give a
few examples:
\begin{center}
\begin{tikzpicture}[scale=0.5]
 \draw[blue,dashed,line width=0.2ex] (0,0)--(6,0)--(6,6)--(0,6)--cycle;
 \foreach \xpos in {1,3,5}
 \foreach \ypos in {0,6}
   \draw[black,fill] (\xpos,\ypos) circle(1ex);
 \foreach \xpos in {0,6}
 \foreach \ypos in {1,3,5}
   \draw[black,fill] (\xpos,\ypos) circle(1ex);
 \draw[black,line width=0.5ex] (0,5)--(6,3);
 \draw[black,line width=0.5ex] (1,0)--(5,6);
 \draw[black,line width=0.5ex] (0,1) arc(-90:90:1cm);
 \draw[black,line width=0.5ex] (3,0) arc(180:0:1cm);

 \begin{scope}[xshift=10cm]
 \draw[blue,dashed,line width=0.2ex] (0,0)--(6,0)--(6,6)--(0,6)--cycle;
 \foreach \xpos in {1,3,5}
 \foreach \ypos in {0,6}
   \draw[black,fill] (\xpos,\ypos) circle(1ex);
 \foreach \xpos in {0,6}
 \foreach \ypos in {1,3,5}
   \draw[black,fill] (\xpos,\ypos) circle(1ex);
 \draw[black,line width=0.5ex] (0,5)--(6,3);
 \draw[black,line width=0.5ex] (1,0)--(5,6);
 \draw[black,line width=0.5ex] (0,3)--(3.5,3.75);
 \draw[black,line width=0.5ex] (3,0) arc(180:0:1cm);
 \end{scope}

 \begin{scope}[xshift=20cm]
 \draw[blue,dashed,line width=0.2ex] (0,0)--(6,0)--(6,6)--(0,6)--cycle;
 \foreach \xpos in {1,3,5}
 \foreach \ypos in {0,6}
   \draw[black,fill] (\xpos,\ypos) circle(1ex);
 \foreach \xpos in {0,6}
 \foreach \ypos in {1,3,5}
   \draw[black,fill] (\xpos,\ypos) circle(1ex);
 \draw[black,line width=0.5ex] (0,5)--(6,3);
 \draw[black,line width=0.5ex] (1,0)--(5,6);
 \draw[black,line width=0.5ex] (0,3)--(3.5,3.75);
 \draw[black,line width=0.5ex] (5,0)--(3.5,3.75);
 \draw[black,line width=0.5ex] (1,6) arc(180:360:1cm);
 \end{scope}
\end{tikzpicture}
\end{center}
Now place a pair of points slightly shifted on either side of each of
the $N$ terminals.  Draw $N$ edges between these $2N$ points by
``turning around'' the hyperedges and isolated vertices of the
hypergraph. We shall refer to this as the surrounding graph. For each
of the above examples this produces:
\begin{center}
\begin{tikzpicture}[scale=0.5]
 \draw[blue,dashed,line width=0.2ex] (0,0)--(6,0)--(6,6)--(0,6)--cycle;
 \foreach \xpos in {0.7,1.3,2.7,3.3,4.7,5.3}
 \foreach \ypos in {0,6}
   \draw[red,fill] (\xpos,\ypos) circle(0.5ex);
 \foreach \xpos in {0,6}
 \foreach \ypos in {0.7,1.3,2.7,3.3,4.7,5.3}
   \draw[red,fill] (\xpos,\ypos) circle(0.5ex);
 \draw[black!30,line width=0.5ex] (0,5)--(6,3);
 \draw[black!30,line width=0.5ex] (1,0)--(5,6);
 \draw[black!30,line width=0.5ex] (0,1) arc(-90:90:1cm);
 \draw[black!30,line width=0.5ex] (3,0) arc(180:0:1cm);
 \draw[red,line width=0.3ex] (0,0.7) arc(-90:90:1.3cm);
 \draw[red,line width=0.3ex] (0,1.3) arc(-90:90:0.7cm);
 \draw[red,line width=0.3ex] (0.7,0)--(2.75,3.15) arc(-30:80:5mm)--(0,4.7);
 \draw[red,line width=0.3ex] (0,5.3)--(3,4.3) arc(260:330:7mm)--(4.7,6);
 \draw[red,line width=0.3ex] (0.7,6) arc(180:360:0.3cm);
 \draw[red,line width=0.3ex] (2.7,6) arc(180:360:0.3cm);
 \draw[red,line width=0.3ex] (5.3,6)--(4.2,4.3) arc(150:260:3mm)--(6,3.3);
 \draw[red,line width=0.3ex] (6,5.3) arc(90:270:0.3cm);
 \draw[red,line width=0.3ex] (6,2.7)--(4,3.4) arc(70:130:5mm)--(1.3,0);
 \draw[red,line width=0.3ex] (6,1.3) arc(90:270:0.3cm);
 \draw[red,line width=0.3ex] (5.3,0) arc(0:180:1.3cm);
 \draw[red,line width=0.3ex] (4.7,0) arc(0:180:0.7cm);

 \begin{scope}[xshift=10cm]
 \draw[blue,dashed,line width=0.2ex] (0,0)--(6,0)--(6,6)--(0,6)--cycle;
 \foreach \xpos in {0.7,1.3,2.7,3.3,4.7,5.3}
 \foreach \ypos in {0,6}
   \draw[red,fill] (\xpos,\ypos) circle(0.5ex);
 \foreach \xpos in {0,6}
 \foreach \ypos in {0.7,1.3,2.7,3.3,4.7,5.3}
   \draw[red,fill] (\xpos,\ypos) circle(0.5ex);
 \draw[black!30,line width=0.5ex] (0,5)--(6,3);
 \draw[black!30,line width=0.5ex] (1,0)--(5,6);
 \draw[black!30,line width=0.5ex] (0,3)--(3.5,3.75);
 \draw[black!30,line width=0.5ex] (3,0) arc(180:0:1cm);
 \draw[red,line width=0.3ex] (0,0.7) arc(-90:90:0.3cm);
 \draw[red,line width=0.3ex] (0.7,0)--(2.5,2.7) arc(-30:120:3.5mm)--(0,2.7);
 \draw[red,line width=0.3ex] (0,3.3)--(2,3.7) arc(-80:80:2mm)--(0,4.7);
 \draw[red,line width=0.3ex] (0,5.3)--(3,4.3) arc(260:330:7mm)--(4.7,6);
 \draw[red,line width=0.3ex] (0.7,6) arc(180:360:0.3cm);
 \draw[red,line width=0.3ex] (2.7,6) arc(180:360:0.3cm);
 \draw[red,line width=0.3ex] (5.3,6)--(4.2,4.3) arc(150:260:3mm)--(6,3.3);
 \draw[red,line width=0.3ex] (6,5.3) arc(90:270:0.3cm);
 \draw[red,line width=0.3ex] (6,2.7)--(4,3.4) arc(70:130:5mm)--(1.3,0);
 \draw[red,line width=0.3ex] (6,1.3) arc(90:270:0.3cm);
 \draw[red,line width=0.3ex] (5.3,0) arc(0:180:1.3cm);
 \draw[red,line width=0.3ex] (4.7,0) arc(0:180:0.7cm);
 \end{scope}

 \begin{scope}[xshift=20cm]
 \draw[blue,dashed,line width=0.2ex] (0,0)--(6,0)--(6,6)--(0,6)--cycle;
 \foreach \xpos in {0.7,1.3,2.7,3.3,4.7,5.3}
 \foreach \ypos in {0,6}
   \draw[red,fill] (\xpos,\ypos) circle(0.5ex);
 \foreach \xpos in {0,6}
 \foreach \ypos in {0.7,1.3,2.7,3.3,4.7,5.3}
   \draw[red,fill] (\xpos,\ypos) circle(0.5ex);
 \draw[black!30,line width=0.5ex] (0,5)--(6,3);
 \draw[black!30,line width=0.5ex] (1,0)--(5,6);
 \draw[black!30,line width=0.5ex] (0,3)--(3.5,3.75);
 \draw[black!30,line width=0.5ex] (5,0)--(3.5,3.75);
 \draw[black!30,line width=0.5ex] (1,6) arc(200:340:10.7mm);
 \draw[red,line width=0.3ex] (0,0.7) arc(-90:90:0.3cm);
 \draw[red,line width=0.3ex] (0.7,0)--(2.5,2.7) arc(-30:120:3.5mm)--(0,2.7);
 \draw[red,line width=0.3ex] (0,3.3)--(2,3.7) arc(-80:80:2mm)--(0,4.7);
 \draw[red,line width=0.3ex] (0,5.3)--(3,4.3) arc(260:330:7mm)--(4.7,6);
 \draw[red,line width=0.3ex] (0.7,6) arc(200:340:14mm);
 \draw[red,line width=0.3ex] (1.3,6) arc(200:340:7.5mm);
 \draw[red,line width=0.3ex] (5.3,6)--(4.2,4.3) arc(150:260:3mm)--(6,3.3);
 \draw[red,line width=0.3ex] (6,5.3) arc(90:270:0.3cm);
 \draw[red,line width=0.3ex] (6,2.7)--(4.3,3.3) arc(70:210:2mm)--(5.3,0);
 \draw[red,line width=0.3ex] (6,1.3) arc(90:270:0.3cm);
 \draw[red,line width=0.3ex] (4.7,0)--(3.7,2.6) arc(10:150:3mm)--(1.3,0);
 \draw[red,line width=0.3ex] (3.3,0) arc(0:180:0.3cm);
 \end{scope}
\end{tikzpicture}
\end{center}

The embedding of $B$ is defined by identifying points on opposing
sides of the frame (to produce the twisted embeddings we further shift
the points on one of the sides cyclically before imposing the
identification).  Let $\ell$ be the number of loops in the surrounding
graph. The partition is of the 1D type if and only if one or more of
these loops is non-homotopic to a point. To determine whether this is
the case it suffices to ``follow'' each loop until one comes back to
the starting point, and determine whether the total signed
displacement in the $x$ and $y$ directions is non-zero.%
\footnote{For the straight embedding one can more simply determine
  whether the signed winding number with respect to any of the two
  periodic directions is non-zero.}
Using this method one sees that the middle partition in the above
three examples is of the 1D type.

If all loops on the surrounding graph have trivial homotopy, one
can use the Euler relation to determine whether the partition is
of the 0D or 2D type. Namely, let $E$ be the sum of all degrees
of the hyperedges in the hypergraph; let $C$ (resp.\ $V$) be the
number of connected components (resp.\ vertices) in the hypergraph
after the identification of opposing sides. Then the combination
\begin{equation}
 \chi = E + 2 C - V - \ell
\end{equation}
equals 0 (resp.\ 2) if the partition is of the 0D (resp.\ 2D) type.

For instance, for the leftmost example we have $E = 3 + 1 + 1 = 5$, $C
= 1$, $V = 6$, and $\ell = 1$, whence $\chi = 0$. And for the
rightmost example one finds $E = 5 + 1 = 6$, $C = 2$, $V = 6$, and
$\ell = 2$, whence $\chi = 2$.

\section{Bond percolation}
\label{sec:bond}

In this section we present our results for bond percolation. The
actual critical polynomials are very large polynomials of degree up to 243
with very large integer coefficients (more than 40 digits), and thus
it does not seem reasonable to make them appear in print. As a
compromise, all the polynomials are collected in the text
file {\tt SC12.m} which is available in electronic form as
supplementary material to this paper.%
\footnote{This file can be processed by {\sc Mathematica} or---maybe
  after minor changes of formatting---by any symbolic computer algebra
  program of the reader's liking.}
The printed version contains only the relevant zeros $p_c \in [0,1]$,
rounded to 15 digit numerical precision.

\subsection{Kagome lattice}

The bond percolation threshold of the kagome lattice is perhaps the
most studied of the unknown bond critical probabilities. Non-rigourous conjectures \cite{Tsallis,Wu79} and
approximations \cite{ScullardZiff06} have appeared in the
literature, as well as rigourous bounds \cite{Wierman2003} and confidence intervals \cite{RiordanWalters07}. To compute polynomials on the kagome lattice, we
considered two families of bases: square (see
section~\ref{sec:sq_bases}) and hexagonal (see
section~\ref{sec:hex_bases}).

\subsubsection{Square bases}

The $n \times n$ square bases with straight and twisted embeddings are
shown in Figure~\ref{fig:squarekagome}. They contain $3 n^2$ vertices
and $6 n^2$ edges.  The percolation thresholds $p_c$ obtained for $n
\le 4$ and twist $k \le \lfloor n/2 \rfloor$ are given in
Table~\ref{tab:squarekagome}. Note that the results for $(n,k) =
(2,0)$ and $(2,1)$ are identical for this lattice; but otherwise the
critical polynomial does depend on $k$.

For the largest ($n=4$) basis, containing $96$ edges, the results for
$p_c$ with the three possible twists have the same first $8$ digits,
perhaps suggesting that at least the first $7$ are actually
correct. By comparing the entries, it also appears that, at least for
$n<4$, the thresholds are correct to the first $n+3$ digits. The
numerical results of Feng, Deng, and Bl\"ote \cite{FengDengBlote08}
place the bond threshold at $p_c=0.524\,404\,99(2)$ using a transfer
matrix approach, and $p_c=0.524\,405\,03(5)$ with Monte Carlo. Our value
is within the error of their second result and can hardly be
considered definitively ruled out by their first. Of course, we cannot
hope that our result is exact, because, as shown in
\cite{Scullard11-2}, no basis of finite size will ever yield the exact
answer. 

\begin{table}
\begin{center}
\begin{tabular}{c|c|c}
 $n$ & twist & $p_c$ \\
\hline \hline
1 & 0 & 0.524\,429\,717\,521\,275 \\
\hline
2 & 0 & 0.524\,406\,723\,188\,232 \\
\ & 1 & 0.524\,406\,723\,188\,232 \\
\hline
3 & 0 & 0.524\,405\,172\,713\,770 \\
\ & 1 & 0.524\,405\,153\,253\,058 \\
\hline
4 & 0 & 0.524\,405\,027\,427\,415 \\
\ & 1 & 0.524\,405\,026\,221\,984 \\
\ & 2 & 0.524\,405\,020\,086\,919 \\
\end{tabular}
\caption{Bond percolation predictions for the kagome lattice on the $n
  \times n$ square bases with various twists.}
\label{tab:squarekagome}
\end{center}
\end{table}

\subsubsection{Hexagonal bases}

The hexagonal bases of size $n$ are shown in
Figure~\ref{fig:hexkagome}. They contain $9 n^2$ vertices and $18 n^2$
edges. As discussed in section~\ref{sec:hex_bases} these bases better
respect the rotational symmetry of the lattice, and hence we expect
the results to be more precise than those with the square bases for a
given number of edges.  Results for $n \le 3$~\footnote{For $n=3$, the basis has $18$ terminals and a very large calculation is necessary. This was done in parallel on Lawrence Livermore National Laboratory's Cab supercomputer, utilizing $2046$ processors, each $2.6$ GHz, for about $20$ hours. The parallel algorithm distributes the state vector over the processors so the primary programming challenge is to ensure that the data is communicated between tasks correctly upon application of the ${\sf B}$, ${\sf L}$ and ${\sf R}$ operators.} are given in Table~\ref{tab:hexkagome}.

We also note that
our $p_c$ for $n \times n$ square bases (\ref{tab:squarekagome}) are monotonically
decreasing with $n$, while those with hexagonal bases are increasing. If these trends hold as $n \rightarrow \infty$, then the kagome threshold satisfies
\begin{equation}
0.524\,404\,998\,266\,288 < p_c < 0.524\,405\,020\,086\,919 \ . \label{eq:kagomebound}
\end{equation}
While this is much more stringent than Wierman's bounds \cite{Wierman2003},
\begin{equation}
 0.5209 < p_c < 0.5291
\end{equation}
his result is completely rigourous while ours is only a guess based on the observed monotonicity in the estimates with $n$. In fact, as we will soon see, the $(3,12^2)$ lattice violates this monotonicity for the $n=3$ hexagonal basis, making (\ref{eq:kagomebound}) even less certain. Nevertheless, the kagome $n=2$ and $3$ (i.e., $72$ and $162$ edges) predictions appear to be converged to at least seven digits, and agree with the transfer matrix result $p_c=0.524\,404\,99(2)$ of Feng, Deng, and
Bl\"ote \cite{FengDengBlote08} to eight decimal places (the limit of their accuracy). Thus we can cautiously conclude that the true bond threshold is
\begin{equation}
 p_c = 0.524\,405\,00(1) \,, \qquad \mbox{Kagome lattice.}
\end{equation}
 More recently, Ding et.~al.~\cite{Ding10} reported
$p_c=0.524\,404\,978(5)$; our results and those of \cite{FengDengBlote08} seem to agree that the error bar of these authors might be slightly underestimated.

\begin{table}
\begin{center}
\begin{tabular}{c|c}
 $n$ & $p_c$ \\
\hline \hline
1 & 0.524\,403\,641\,312\,579 \\
\hline
2 & 0.524\,404\,993\,638\,028 \\
\hline
3 & 0.524\,404\,998\,266\,288
\end{tabular}
\caption{Bond percolation predictions for the kagome lattice on the $n$--sided hexagonal bases.}
\label{tab:hexkagome}
\end{center}
\end{table}

\subsection{$(4,8^2)$ lattice}

\begin{figure}
\begin{center}
\includegraphics{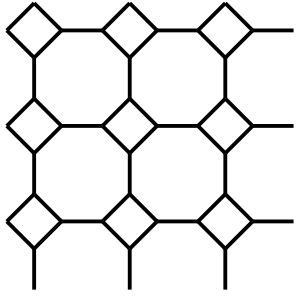}
\caption{The $3 \times 3$ square basis, with unspecified embedding, for the $(4,8^2)$ lattice.}
\label{fig:FE3x3}
\end{center}
\end{figure}

We computed the critical polynomials for the $n \times n$ square bases
on the $(4,8^2)$ lattice (see Figure~\ref{fig:FE3x3}). As this graph
does not have the kagome lattice's hexagonal symmetry, there are no
corresponding hexagonal bases.  Results for $n \le 4$ are given in
Table~\ref{tab:foureight}, with the twists $k \le \lfloor n/2 \rfloor$
defined identically to the kagome case.  Note that the cases $(n,k) =
(2,0)$ and $(2,1)$ now produce different results.

The bond threshold of this lattice has not been studied as thoroughly
as that of the kagome lattice, and apparently the only high-precision
result is Parviainen's \cite{Parviainen}, $p_c=0.676\,802\,32(63)$. Our $4 \times 4$ results are within two standard deviations.

\begin{table}
\begin{center}
\begin{tabular}{c|c|c}
 $n$ & twist & $p_c$ \\
\hline \hline
1 & 0 & 0.676\,835\,198\,816\,406 \\
\hline
2 & 0 & 0.676\,811\,051\,133\,795 \\
\ & 1 & 0.676\,805\,751\,049\,826 \\
\hline
3 & 0 & 0.676\,805\,010\,886\,365 \\
\ & 1 & 0.676\,803\,989\,559\,125 \\
\hline
4 & 0 & 0.676\,803\,693\,656\,055 \\
\ & 1 & 0.676\,803\,476\,910\,363 \\
\ & 2 & 0.676\,803\,329\,691\,626 \\
\end{tabular}
\caption{Bond percolation predictions for the $(4,8^2)$ lattice on the $n \times n$ square bases with various twists.}
\label{tab:foureight}
\end{center}
\end{table}

\subsection{$(3,12^2)$ lattice}

The $(3,12^2)$ lattice bears more than a passing resemblance to the
kagome lattice. Employing the analogous $n \times n$ square bases and
twists, we find the results in Table~\ref{tab:squareTT}.

\begin{table}
\begin{center}
\begin{tabular}{c|c|c}
 $n$ & twist & $p_c$ \\
\hline \hline
1 & 0 & 0.740\,423\,317\,919\,897 \\
\hline
2 & 0 & 0.740\,420\,992\,429\,996 \\
\ & 1 & 0.740\,420\,992\,429\,996 \\
\hline
3 & 0 & 0.740\,420\,818\,821\,979 \\
\ & 1 & 0.740\,420\,817\,594\,340 \\
\hline
4 & 0 & 0.740\,420\,802\,130\,112 \\
\ & 1 & 0.740\,420\,802\,158\,172 \\
\ & 2 & 0.740\,420\,801\,695\,085 \\
\end{tabular}
\caption{Bond percolation predictions for the $(3,12^2)$ lattice on the $n \times n$ square bases with various twists.}
\label{tab:squareTT}
\end{center}
\end{table}

Like the kagome lattice, the bond threshold on this lattice has
been studied extensively. Parviainen, using simulations, gives the
threshold as $p_c=0.740\,421\,95(80)$. More recent transfer matrix
work by Ding et~al.~\cite{Ding10} gives $p_c=0.740\,420\,77(2)$,
whereas Ziff and Gu \cite{ZiffGu09} report $p_c=0.740\,420\,81(10)$
based on a fitting method.

\begin{table}
\begin{center}
\begin{tabular}{c|c}
 $n$ & $p_c$ \\
\hline \hline
1 & 0.740\,420\,702\,159\,477 \\
\hline
2 & 0.740\,420\,799\,397\,205 \\
\hline
3 & 0.740\,420\,798\,850\,745 \\
\end{tabular}
\caption{Bond percolation predictions for the $(3,12^2)$ lattice on the hexagonal bases of side $n$.}
\label{tab:hexTT}
\end{center}
\end{table}

Results with the hexagonal basis are shown in
Table~\ref{tab:hexTT}\footnote{The $n=3$ calculation required 20 hours on 4092 processors, each 2.6 GHz.}. While the square basis values seem to approach the exact solution from above, as in the kagome case, the hexagonal bases deviate from the trend of approach from below with the $n=3$ result. Nevertheless, it is this latter estimate that we expect to be the most accurate, and we cautiously conclude that the true bond threshold of $(3,12^2)$ is
\begin{equation}
 p_c = 0.740\,420\,800(2) \,, \qquad \mbox{$(3,12^2)$ lattice.}
\end{equation}
This value is one order of magnitude more precise than the most recent
numerical work and demonstrates the potential of the critical
polynomials for determining high-precision critical thresholds.

\section{Site percolation}

The condition (\ref{eq:2D0D}) allows for a straightforward extension
to site percolation. We first demonstrate that the results for the
smallest possible bases correctly retrieve the thresholds for exactly
solvable lattices. We then present results with large bases for the
square and hexagonal lattices.

\subsection{Exactly solvable lattices}
In some sense, site percolation is more fundamental than the bond problem. This is because every lattice has a line graph, or covering lattice, which maps bond percolation to a corresponding site problem. The covering lattice of $L$ is formed by placing a vertex on every edge of $L$, and drawing edges between vertices that cover adjacent edges of $L$. The resulting graph is usually not planar, but it is obvious that site percolation on the covering lattice is identical to bond percolation on $L$. The inverse construction is rarely possible. That is, not every site problem can be mapped to a bond problem (without resorting to hyperedges) on an underlying $L$, and in this sense bond percolation is a special case of the site problem. Although there are now lattices for which the site thresholds are known that are neither self-matching nor the covering lattices of bond problems \cite{Scullard06, Ziff06}, among the Archimedean lattices only the triangular, which is self-matching and thus has $p_c^{\mathrm{site}}=1/2$, and the kagome and $(3,12^2)$ lattices, which are the line graphs of the hexagonal and doubled-bond hexagonal lattices respectively, have known site thresholds. Here, we apply the method to these solvable cases to verify that the condition (\ref{eq:2D0D}) does reproduce exact solutions.

\subsubsection{Triangular lattice}

Site percolation configurations on the triangular lattice can be conveniently described as colourings of the faces on the dual, hexagonal lattice.
The simplest possible basis $B$ consists of just a single hexagon, for which we use the hexagonal embedding. Clearly $P(2D;B) = v$ and
$P(0D;B) = 1$, so that (\ref{eq:2D0D}) yields $v_c = 1$ or $p_c = 1/2$, which is indeed the correct answer.

\subsubsection{Kagome and $(3,12^2)$ lattices}

For the kagome lattice we similarly consider face colourings of the dual, diced lattice, which is a tiling of the plane with three differently oriented
lozenges. The simplest basis $B$ consists of three different lozenges inscribed in a hexagon. We have then $P(2D;B) = v^3$,
$P(1D;B) = 3 v^2$, and $P(0D;B) = 3 v + 1$. Application of (\ref{eq:2D0D}) then gives the critical polynomial
\begin{equation}
 1 - 3 p^2 + p^3 = 0 \,,
\end{equation}
and the relevant zero $p_c = 1 - 2 \sin(\pi/18) = 0.652\,704 \cdots$ provides the exactly known threshold.

A very similar computation for the $(3,12^2)$ lattice, using a basis of six sites, gives the same answer as for the kagome lattice, except that $p$ is replaced by $p^2$.

\subsection{Square lattice}
For the square lattice we use rectangular bases of $n \times m$ sites
(see section~\ref{sec:sq_bases}). The site polynomials on this lattice
are not found any more efficiently with the transfer matrix of
section~\ref{sec:tm} than by simply using the brute force approach of
generating all $2^{n m}$ configurations and directly computing the
probabilities $P(2D)$ and $P(0D)$. Therefore we take the latter
approach in this case\footnote{Polynomials for bases up to $6 \times 5$ could be computed
  on an ordinary desktop. To get the $6 \times 6$ result, we use
  $2048$ processors, each $2.4$ GHz, on Lawrence Livermore National
  Laboratory's Atlas supercomputer. In contrast to the transfer matrix, the parallel implementation is somewhat trivial as it is
  effected by simply dividing the $2^{36}$ configurations over the processors
  so that each one handles $2^{25}$ with little inter-processor communication required. The calculation completes
  in about an hour.}. The results for $n,m \le 6$ are shown in
Table~\ref{tab:squaresite}. 

\begin{table}
\begin{center}
\begin{tabular}{c|c}
 basis & $p_c$ \\
\hline \hline
$1 \times 1$ & 0.5 \\
$2 \times 2$ & 0.541\,196\,100\,146\,197 \\
$3 \times 3$ & 0.586\,511\,455\,112\,676 \\
$3 \times 4$ & 0.588\,361\,985\,284\,352 \\
$4 \times 4$ & 0.590\,672\,112\,331\,028 \\
$4 \times 5$ & 0.591\,269\,973\,846\,402 \\
$5 \times 5$ & 0.591\,988\,256\,518\,334 \\
$5 \times 6$ & 0.592\,167\,665\,055\,742 \\
$6 \times 6$ & 0.592\,395\,070\,817\,704 \\
\end{tabular}
\caption{Site percolation predictions for the square lattice on $n \times m$ rectangular bases.}
\label{tab:squaresite}
\end{center}
\end{table}

The site threshold on the square lattice is the subject of perhaps the most numerical studies of all the Archimedean percolation problems \cite{FengDengBlote08,Ziff92,Newman2000,Ziff2002,Deng05,Lee07,Lee08}. To take the most recent of these, Lee \cite{Lee08} found $p_c= 0.592\,745\,98(4)$ by a Monte Carlo scheme, whereas Feng, Deng and Bl\"{o}te \cite{FengDengBlote08} used both Monte Carlo, $p_c=0.592\,746\,06(5)$, and transfer matrix, $p_c=0.592\,746\,05(3)$, methods. These results are all within each other's error bars and unanimously and decisively rule out our best polynomial prediction. Compared with the bond percolation results
presented here, it is striking how poorly the polynomials perform for
this problem. Even for the $36^{\mathrm{th}}$-order polynomial of the
$6 \times 6$ basis, we are left with a prediction that is barely
within $3.5 \times 10^{-4}$ of the numerical answer, whereas a
polynomial for a bond problem is typically off by only $10^{-7}$ at
this order \cite{Scullard12,Scullard11-2}.

\subsection{Hexagonal lattice}
Although the bond percolation threshold for the hexagonal lattice has been known rigourously for a long time \cite{Wierman81}, and conjecturally for even longer \cite{SykesEssam}, its exact site threshold remains elusive. Before accurate numerical results were available, it was guessed, based on a star-triangle argument, that the site threshold is given by $p_c=1/\sqrt{2} \approx 0.707107$ \cite{Kondor80}. Although this is now known to be incorrect, it is reasonably close and in fact, the critical polynomial for the two-site basis also makes this prediction\footnote{Interestingly, this is the {\it exact} site threshold for a different lattice, the martini-A \cite{Scullard06}, which bears some resemblance to the hexagonal lattice}. We improve upon this estimate by employing the hexagonal bases of Figure \ref{fig:hexkagome} with each triangle of the kagome lattice replaced with a site (the kagome lattice is the medial graph of the hexagonal) and the transfer operators {\sf B}, {\sf L} and {\sf R} given by (\ref{eq:hexB})--(\ref{eq:hexR}). 

Predictions for $n=1$, $2$ and $3$ 
(for $n=3$, a parallel computation was necessary, utilizing $2046$ processors, which completed in about three hours), are roots of $6^{\mathrm{th}}$, $24^{\mathrm{th}}$, and $54^{\mathrm{th}}$ order polynomials. These thresholds are presented in Table \ref{tab:hexsite}. Suding and Ziff's Monte Carlo estimate \cite{SudingZiff99} places the critical probability around $p_c=0.697\,043(3)$. A more recent transfer matrix result of Feng, Deng, and Bl\"{o}te \cite{FengDengBlote08} is $p_c=0.6970402(1)$, and, although it is within $2.6 \times 10^{-6}$, our $n=3$ prediction is clearly ruled out. This is in sharp contrast to the bond results for the kagome and $(3,12^2)$ lattices, which already challenge the numerical results at $n=2$. However, this is still better than the situation for site percolation on the square lattice. The $n=2$ hexagonal basis contains $24$ sites and makes a prediction within $2.2 \times 10^{-5}$ of the numerical value, which is an order of magnitude better than the $36$--site square lattice basis. We will have more to say about this below.
\begin{table}
\begin{center}
\begin{tabular}{c|c}
 $n$ & $p_c$ \\
\hline \hline
$1$ & 0.691\,538\,728\,617\,958 \\
\hline
$2$ & 0.697\,018\,214\,522\,145 \\
\hline
$3$ & 0.697\,037\,409\,746\,762 \\
\end{tabular}
\caption{Site percolation predictions for the hexagonal lattice on hexagonal bases of side length $n$.}
\label{tab:hexsite}
\end{center}
\end{table}

\section{Discussion}
In this work, we have given a re-definition, equation (\ref{eq:2D0D}), of the generalised critical polynomial which was defined previously through contraction-deletion. While the old definition placed a practical limit on the computation of polynomials of about $36^{\mathrm{th}}$ order, this new definition allowed us to use a transfer matrix approach to calculate polynomials up to degree $243$. The results presented here provide very clear evidence for the conjecture, put forward, for example, in \cite{Scullard11} and \cite{Jacobsen12}, that the root in $[0,1]$ of a generalised critical polynomial, $P_B(p)$, provides either the exact percolation threshold, or gives an approximation that approaches the exact answer in the limit of an appropriately infinite basis $B_\infty$. Specifically, it was conjectured in \cite{Jacobsen12} that, as long as the aspect ratio of the limiting $B_\infty$ is non-zero and finite, then all possible $B_\infty$ make the same prediction for the critical probability. We have provided evidence for this as well, through the use of both square and hexagonal bases for the kagome and $(3,12^2)$ lattices.

Needless to say, there is a fair degree of conjecture involved in this work. First of all, the equivalence between the contraction-deletion definition and the probabilistic definition (\ref{eq:2D0D}) of the polynomials, which we found essentially by inspection, needs to be firmly established. Furthermore, the central idea behind all our computations, namely that (\ref{eq:2D0D}) fixes the critical point in the scaling limit, follows from universality and so is possibly very difficult to prove in general. Even granted universality, it is not clear why this toroidal crossing probability should be the one that provides the most rapid passage into the scaling limit, at least as far as the critical threshold is concerned. Nevertheless, all these things appear to be true, as we hope we have demonstrated, and, even absent the wanted rigour, this method produces very accurate thresholds and may even come to supplant other numerical techniques for determining critical probabilities, at least for bond problems.

The kagome and $(3,12^2)$ bond results seemingly cannot be ruled out by current numerics, but the square site and hexagonal predictions are not as competitive. The method seems to perform best for families of bases in which the ratio, which we denote $\zeta(n)$, of the number of boundary vertices, or terminals, to the number of internal elements (sites or bonds) is large. The hexagonal bases of side $n$ have $6n$ terminals, but the number of interior elements depends on the lattice chosen. For the hexagonal site problem, there are $6 n^2$ sites so $\zeta(n)=1/n$, for kagome bond percolation $\zeta(n)=1/(3 n)$, while for $(3,12^2)$ $\zeta(n)=2/(9 n)$. Even at $n=2$, the latter two problems make predictions comparable to numerics, whereas the $n=3$ hexagonal site prediction is ruled out, and it is tempting to believe that the speed with which the estimates approach the exact answer is related to the speed with with $\zeta(n)$ goes to $0$ as $n \rightarrow \infty$. Further support for this is found by considering the square site problem, in which the square bases have $4n$ terminals and $n^2$ sites, or $\zeta(n)=4/n$, so the worst estimates are given by the system with the slowest convergence of $\zeta(n)$ to $0$.

There are many other directions for future work. The condition (\ref{eq:2D0D}) has a generalization to the $q-$state Potts model, allowing predictions of critical points for general $q$ of similar quality to those reported here for $q=1$. This is the subject of ongoing study. Also, the general strategy employed here may be applicable to other lattice models, for which exact results are known only on some lattices. Finally, we mention that the generalised critical polynomial can be defined through contraction-deletion in higher dimensions, but it is not yet clear whether they provide any useful information about the critical point, or whether there is a higher-dimensional equivalent of equation (\ref{eq:2D0D}).

\section*{Acknowledgments}

The work of JLJ was supported by the Agence Nationale de la Recherche
(grant ANR-10-BLAN-0414:~DIME) and the Institut Universitaire de
France. This work was partially (CRS) performed under the auspices of
the U.S. Department of Energy by Lawrence Livermore National
Laboratory under Contract DE-AC52-07NA27344. CRS
wishes to thank Bob Ziff for discussions and collaboration on related
work, and both authors thank Jim Glosli at LLNL for helpful advice on the parallel implementation of the transfer matrix code. We are grateful to the Mathematical Sciences Research
Institute at the University of California, Berkeley for hospitality
during the programme on Random Spatial Processes where this work was
initiated. Finally, CRS thanks the Institute for Pure and Applied Mathematics at UCLA, where part of this work was performed.

\section*{References}
\bibliographystyle{unsrt}
\bibliography{SJ}

\end{document}